\documentclass[journal]{IEEEtran}

\if CLASSOPTIONcompsoc
  \usepackage[nocompress]{cite}
\else
  \usepackage{cite}
\fi

\usepackage[utf8x]{inputenc}
\usepackage{listings}
\usepackage{color}
\usepackage{url}
\usepackage{amsmath}
\usepackage{cite}
\usepackage{graphicx}
\usepackage{color}
\usepackage[bookmarks=false]{hyperref}
\usepackage{amssymb,amsmath,dsfont}
\usepackage{amsthm}
\usepackage[linesnumbered,lined, ruled]{algorithm2e}
\usepackage{lipsum}
\usepackage{bbm}
\usepackage{tablefootnote}

\usepackage{tikz}
\usetikzlibrary{automata,arrows,positioning,calc}

\usepackage{multirow}

\usepackage{breqn}
\SetAlFnt{\footnotesize}

\usepackage{graphics}

\newtheorem{theorem}{Theorem}
\newtheorem{corollary}{Corollary}

\newtheorem{lemma}{Lemma}

\theoremstyle{remark}
\newtheorem*{remark}{Remark}

\theoremstyle{remark}

\newtheorem*{exmp}{Example}

\usepackage{csquotes}
\usepackage{amsmath}
\usepackage{amssymb}
\usepackage{subcaption}
\usepackage{graphicx}
\usepackage{accents}
\usepackage{mathtools}
\usepackage[normalem]{ulem}

\SetCommentSty{mycommfont}

\newcommand{\tolL}{\Delta^{\mathrm{L}}}
\newcommand{\tolU}{\Delta^{\mathrm{U}}}
\newcommand{\tai}{\mathcal{H}_{\mathrm{TAI}}}
\newcommand{\Reals}{\mathbb{R}}

\graphicspath{figures/}

\usepackage{mathtools}
\DeclarePairedDelimiter\floor{\lfloor}{\rfloor}
\DeclarePairedDelimiter\ceil{\lceil}{\rceil}

\SetKwRepeat{Repeat}{repeat}{until}
\SetKwRepeat{Forever}{repeat}{forever}
\SetKwRepeat{On}{on}{end}
\SetNlSty{bfseries}{\color{black}}{}

\newtoggle{PROOF}
\toggletrue{PROOF}
\begin{document}

\title{Analysis of Dampers in Time-Sensitive Networks with Non-ideal Clocks}

% \author{
% \IEEEauthorblockN{Author}
% \IEEEauthorblockA{EPFL \\
% Email: }
% \and
% \IEEEauthorblockN{Author}
% \IEEEauthorblockA{EPFL \\
% 	Email: }
% \and
% \IEEEauthorblockN{Author}
% \IEEEauthorblockA{EPFL \\
% 	Email: }
% \and
% \IEEEauthorblockN{Author}
% \IEEEauthorblockA{EPFL \\
% 	Email: }
% }

\author{\IEEEauthorblockN{
		Ehsan Mohammadpour, Jean-Yves Le Boudec
		\\}
\IEEEauthorblockA{\'Ecole Polytechnique F\'ed\'erale de Lausanne, Switzerland\thanks{This work was supported by Huawei Technologies Co., Ltd. in the framework of the project Large Scale Deterministic Network. The authors thank Bingyang Liu and Shoushou Ren for fruitful discussions. }\\
$\{$firstname.lastname$\}$@epfl.ch}}

\maketitle
%\copyrightnotice
\begin{abstract}
Dampers are devices that reduce delay jitter in the context of time-sensitive networks, by delaying packets for the amount written in packet headers. 
% why important ?
Jitter reduction is required by some real-time applications; beyond this, dampers have the potential to solve the burstiness cascade problem of deterministic networks in a scalable way, as they can be stateless. 
% why is it a problem ?
Dampers exist in several variants: some apply only to  earliest-deadline-first schedulers, whereas others can be associated with any packet schedulers; some enforce FIFO ordering whereas some others do not. Existing analyses of dampers are specific to some implementations and some network configurations; also, they assume ideal, non-realistic clocks. In this paper, we provide a taxonomy of all existing dampers in general network settings and analyze their timing properties in presence of non-ideal clocks. In particular, we give formulas for computing residual jitter bounds of networks with dampers of any kind. We show that non-FIFO dampers may cause reordering due to clock non-idealities and that the combination of FIFO dampers with non-FIFO network elements may very negatively affect the performance bounds. Our results can be used to analyze timing properties and burstiness increase in any time-sensitive network, as we illustrate on an industrial case-study.
%keywords: Damper, Jitter, Time-sensitive network, Burstiness cascade, non-ideal clocks, Network Calculus
\end{abstract}

\IEEEpeerreviewmaketitle

\section{Introduction}\label{sec:intro}
Time-sensitive networks provide guarantees for applications in the automobile, automation, space, avionics and video industries \cite{ieeeDraftStandardLocal2019b,iecIECIEEE608022019,ecssSpaceWireLinksNodes2008,AFDX,TTE,ieeeAVB}. IEEE Time Sensitive Networking (TSN) working group \cite{tsn} and the IETF Deterministic Networking (DetNet) working group \cite{detnet} provide standardization for such networks. The goal of time-sensitive networks is to fulfill flow requirements on worst-case delay and jitter (defined as the difference between worst-case and best-case delays), in-order packet delivery, as well as zero congestion loss and seamless redundancy~\cite{ieeeIEEEStandardLocal2017,rfc8655}. The emergeance of applications with low jitter requirement in large-scale time-sensitive networks, such as industrial Internet of Things \cite{itu-y3000} and electricity distribution\cite{tsn-profile-service-provider}, questions the performance of existing queuing and shaping mechanisms such as Credit-Based Shaper, IEEE 802.1Qch Cyclic Queuing and Forwarding (CQF) \cite{p8021qch}, and Deficit Round Robin \cite{tabatabaee2021deficit}. This issue can be addressed with dampers, which are mechanisms to reduce delay jitter in time-sensitive networks \cite{verma_delay_1991,zhang_rate-controlled_1993,cruz_sced+:_1998}.

A damper delays every time-sensitive packet by an amount written in a packet header field, called damper header, which carries an estimate of the earliness of this packet with respect to a known delay upper-bound of upstream systems. This ideally leads to zero jitter; in practice, there is still some small residual jitter, due to errors in acquiring timestamps and in computing and implementing delays. As a positive side effect, dampers create packet timings that are almost the same as at the source, with small errors due to residual jitter, and thus cancel most of the burstiness increase imposed by the network.
\cite[Lemma 1]{mohammadpour2020packet}. The residual burstiness increase that remains when dampers are used is not influenced by the burstiness of cross-traffic. Thus, dampers solve the burstiness cascade issue \cite{charny_delay_2000}: individual flows that share a resource dedicated to a class may see their burstiness increase, which may in turn increase the burstiness of other downstream flows. Furthermore, dampers are stateless, unlike some TSN shaping mechanisms, e.g., Asynchronous Traffic Shaping (ATS) \cite{ieee8021qcr}. Solving the burstiness cascade in a stateless manner makes the dampers of interest for large-scale time-sensitive networks.

%\todo{what is the problem}
Several implementations of dampers have been proposed; the older ones are associated with specific schedulers such as earliest-deadline-first \cite{verma_delay_1991,cruz_sced+:_1998} and static priority \cite{zhang_rate-controlled_1993}; the recent implementations can coexist with any scheduling mechanism \cite{grigorjewMetzgerHossfeldetal_2020,rgcq,fopleq}. Some of these implementations enforce dampers to behave in a FIFO manner \cite{cruz_sced+:_1998,fopleq,grigorjewMetzgerHossfeldetal_2020} and some do not \cite{zhang_rate-controlled_1993,rgcq}. Analysis of damper is crucial to provide guarantees for applications in the context of time-sensitive networks. In the existing works, \cite{rgcq,fopleq} did not provide any analysis; others analyze only their implementation and under limited assumptions on the network settings. Also, existing analyses assume that the network operates with one ideal clock; in practice, this assumption does not hold and may have non-negligible side effects. Recently, the effect of non-ideal clocks on regulators was analyzed and a clock model was proposed in the context of time-sensitive networks \cite{thomas_time_2020}, which we use in this paper.

%\todo{what does this paper do...}
We first present a taxonomy of dampers that classifies the existing implementations into dampers with or without FIFO constraint. Then, under general network configuration with non-ideal clocks, we provide formulas to compute tight delay and jitter bounds for dampers without FIFO constraint (Theorems~\ref{thm:perhop-concrete-delay} and~\ref{thm:e2e-concrete}); we see that the impact of non-ideal clocks can be non-negligible in cases with low jitter requirements. As a result of this analysis, we derive conditions in which clock synchronization throughout a network does not affect the performance of dampers. Moreover, we capture the propagation of arrival curve at the output of dampers and see how this can solve the burstiness cascade issue. Next, we show that existing implementations of dampers without FIFO constraints may cause undesired packet reordering due to clock non-idealities, even in synchronized networks. This problem is avoided with dampers that enforce FIFO constraints; however, the effect on their timing properties was not analysed in the literature and we bridge this gap in this paper. We model two classes of dampers with FIFO constraint: re-sequencing dampers and head-of-line dampers. For the former class, we show that when all network elements are FIFO, the delay and jitter bounds are not affected by the re-sequencing operation (Theorem \ref{thm:perhop-fifodamper-delay-fifo}). For the latter class, there is a small penalty due to head-of-line queuing, which we quantify exactly (Theorem \ref{thm:perhop-jcats-delay-fifo}).
In contrast, if some network elements are non FIFO, the jitter bounds for dampers with FIFO constraint can be considerably larger (Theorems~\ref{thm:perhop-fifodamper-delay-nonfifo} and~\ref{thm:perhop-jcats-delay-nonfifo}). We finally evaluate our results in an industrial case-study.

The rest of the paper is as follows. Section~\ref{sec:related} presents the state-of-the-art. 
Section~\ref{sec:sys} describes the system model, terminology, clock model and all assumptions. Section~\ref{sec:dm} presents a taxonomy of the existing dampers. The analysis of dampers without FIFO constraint is presented in Section \ref{sec:analysis-noFIFO}. Packet reordering scenarios due to non-ideal clocks are presented in Section \ref{sec:concrete-reordering}. The analysis of dampers with FIFO constraint is given in Section \ref{sec:analysis-FIFO}. Section \ref{sec:eval} provides a numerical evaluation for an industrial case-study and Section \ref{sec:conclusion} concludes the paper.

\section{Related Works}\label{sec:related}

The concept of dampers was introduced by Verma et. al \cite{verma_delay_1991}, under the name \textit{delay-jitter regulator}, in combination with earliest-deadline-first (EDF) scheduling. In this scheme, a per-flow regulator is placed at every node to delay a packet as much as its earliness in the previous node; the earliness is the time difference between the delay that a packet was supposed to experience and the actual delay that is measured by time-stamping. Later, Zhang et. al. \cite{zhang_rate-controlled_1993} proposed Rate-Control Static Priority (RCSP) scheduling to avoid coupling of delay and bandwidth allocation in the EDF schedulers mentioned in \cite{verma_delay_1991}.  We describe RCSP in Section~\ref{sec:dm}.

The term \textit{damper} was first used by Rene Cruz \cite{cruz_sced+:_1998} as a conceptual network element that slows down the traffic passing through it. In \cite{cruz_sced+:_1998}, dampers are used in relationship with SCED (Service Curve Earliest Deadline) scheduling to avoid extra queuing as was proposed by \cite{zhang_rate-controlled_1993}. With this scheme, called SCED+, a flow traverses a few virtual paths (each is a sequence of switches) with guaranteed service curves and damper curves. Then, at the entrance of each virtual path, for every packet of the flow and every switch in the virtual path, initial and terminal eligibility times are computed using the service and damper curves; a packet is released from a switch within its initial and terminal eligibility times.

Recently, a few implementations of damper are proposed that can be used in combination with any scheduling mechanism. Grigorjew et. al. \cite{grigorjewMetzgerHossfeldetal_2020} implement damper as a shaper in relation with Asynchronous Traffic Shaping (ATS), IEEE 802.1 Qcr \cite{ieee8021qcr}; we refer to their scheme as jitter-control ATS. It is assumed in \cite{grigorjewMetzgerHossfeldetal_2020} that the input flows are constrained by leaky-bucket arrival curve and all the elements inside the network, including the switching fabrics, output port queues and the ATS, are FIFO for the packets that share the same queues inside ATS. Rotated gate-control-queues (RGCQ) \cite{rgcq} is an implementation of a damper integrated with the queuing system of a scheduler. Flow-order preserving latency-equalizer (FOPLEQ) \cite{fopleq} is as an extension of RGCQ to preserve the per-flow order of the packets according to its entrance to FOPLEQ. Section \ref{sec:dm} describes the details of these implementations. These previous works do not provide delay analysis or do it in restricted settings. In particular, clock non-idealities are ignored. In~\cite{thomas_time_2020} clock non-idealities are modelled in the context of time-sensitive networks and the impact on timing analyses is explained in detail. In this paper, we apply this clock model to networks with dampers of various kinds.

Dampers can be used to reduce delay jitter and thus to provide end-to-end services with a low jitter guarantee. An alternative method to provide low jitter, Cyclic Queuing and Forwarding (CQF), also known as Peristaltic Shaper, was introduced by IEEE Time-Sensitive Networking (TSN)  \cite{p8021qch}, \cite[Annex T]{ieee8021Q}. According to CQF, for each priority class, there are two cyclic queues; in each cycle, while one queue is being served, the other enqueues the arriving packets. The cycles change periodically and the queues swap their operations with each other. CQF relies on very different mechanisms and assumptions than dampers; its analysis is out of the scope of this paper.
\section{System Model}\label{sec:sys}
 We consider a network that contains a set switches or routers, hosts and links with fixed capacity. Every flow follows a fixed path, has a finite lifetime and emits a finite, but arbitrary, number of packets. We consider unicast flows with known arrival curves at their sources (i.e. there are known bounds on the number of bits or packets that can be emitted by a flow within any period of time).
\subsection{Terminology}\label{sec:term}
We call \emph{jitter-compensated system} (JCS) any delay element or aggregate of delay elements with known delay and jitter bounds, for which we want to compensate jitter by means of dampers. This is typically the queuing system on the output port of a switch or router used in time-sensitive networks. It can also be a switching fabric or an input port processing unit, or even a larger system. For time-sensitive flows, a JCS should be able to time stamp packet arrivals and departures using the available local times. It should also increment the damper header field in every time-sensitive packet (if one is present) by an amount equal to an estimate of the earliness of this packet with respect to the known delay upper-bound at the JCS for the class of traffic that this packet belongs to. If no damper header is present, it inserts one, with a value equal to the estimated earliness.\footnote{We choose this method of carrying earliness in packet headers for ease of presentation. Another method consists in each JCS inserting a separate damper header: a packet then has as many damper headers as JCSs between dampers, and the earliness to be compensated at a damper is the sum of all these values. The discussion of such methods is out of the scope of this paper, as it does not affect the timing analysis presented here.}
 The operation of the damper header update (DHU) unit is described in Section \ref{sec:damper-header}. When a time-sensitive flow crosses a JCS, for actual jitter removal to occur, there must be a downstream damper on the path of the flow. For example, if the JCS is a switch output port, the next downstream damper is typically located on the output port of the next downstream switch.

It is generally not possible, or required, to remove delay jitter in all network elements, because time stamping and DHU come with a cost. Therefore, it is required, for our timing analysis, to consider what we call \emph{bounded-delay systems} (BDSs), defined as any delay element or aggregate of delay elements with known delay and jitter bounds, and for which we do not compensate jitter. Constant delay elements (e.g. an output link propagation delay), variable delay elements with very low jitter (e.g., very high speed backbone network) and other delay elements without DHU unit are examples of BDSs.

A \emph{damper} is a system that delays every time-sensitive packet, using its local clock, for a duration approximately equal to the damper header (if any, else the damper does not delay the packet).
 Such a damper header was inserted/updated in the upstream JCSs between this damper and the previous upstream damper or the source of the flow. The damper also resets the damper header, so that the next downstream damper will see only the earliness accumulated downstream of this damper. Designing a stand-alone damper is a challenge, because such a damper may need to release a large number of packets instantly or within a very short time, which might not be feasible. This is why damper implementations are often associated with queuing systems; then, the time at which a damper releases a packet is simply the time at which the packet becomes visible to the queuing system. We classify and model existing designs of dampers in Section \ref{sec:dm}.

\begin{figure}[h]
	\centering
	\includegraphics[width= \linewidth]{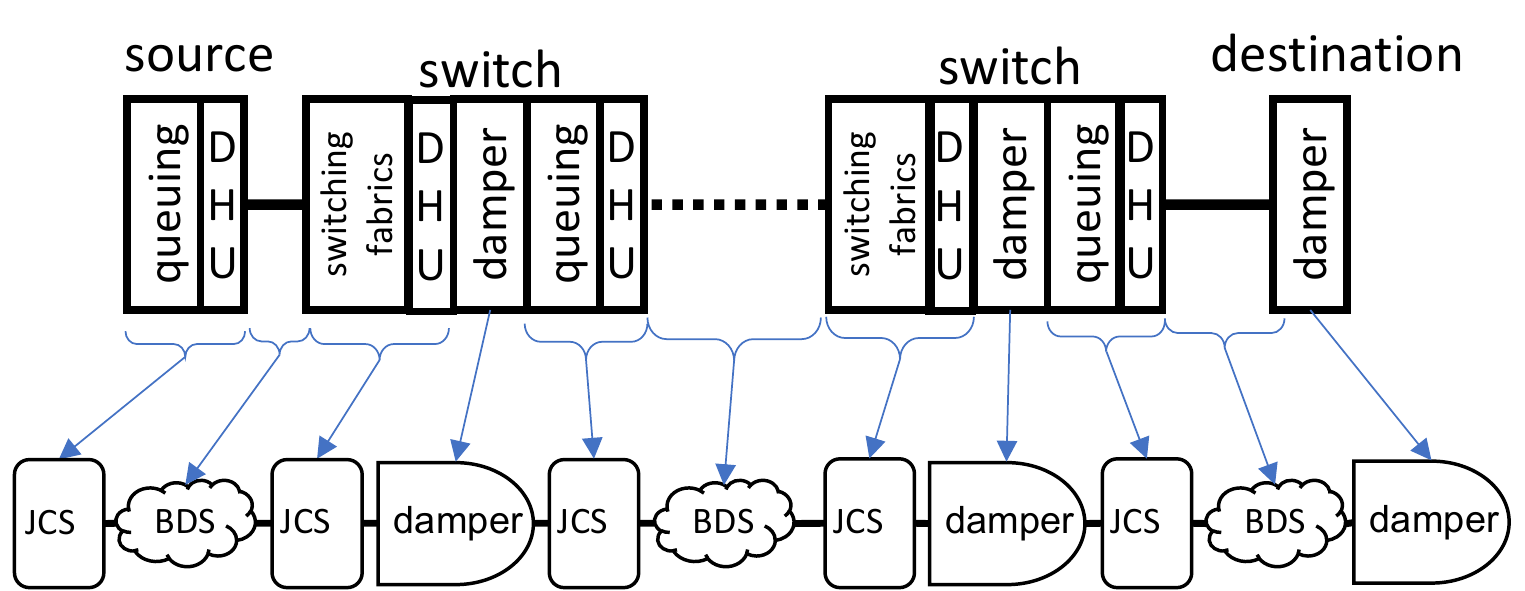}
	\caption{A local-area time-sensitive network example.}
	\label{fig:tsn-usecase}
\end{figure}

\textbf{Example 1.} Fig. \ref{fig:tsn-usecase} shows an example flow path within a local-area time-sensitive network where we want to compensate the jitter imposed by the output queuing systems and switching fabrics by means of dampers. Therefore, for each of the switching fabrics and the queuing systems, a DHU unit is placed to perform the damper header update; finally a damper is placed before each queuing system to remove the imposed jitter by the upstream switching fabric and queuing system. For example, the damper in the first switch compensates the jitter imposed by the queuing system of the source and the switching fabric of the first switch. Note that the propagation delay is constant and seen as a BDS. Here, the different clocks need not to be synchronized.

\begin{figure}[h]
	\centering
	\includegraphics[width= \linewidth]{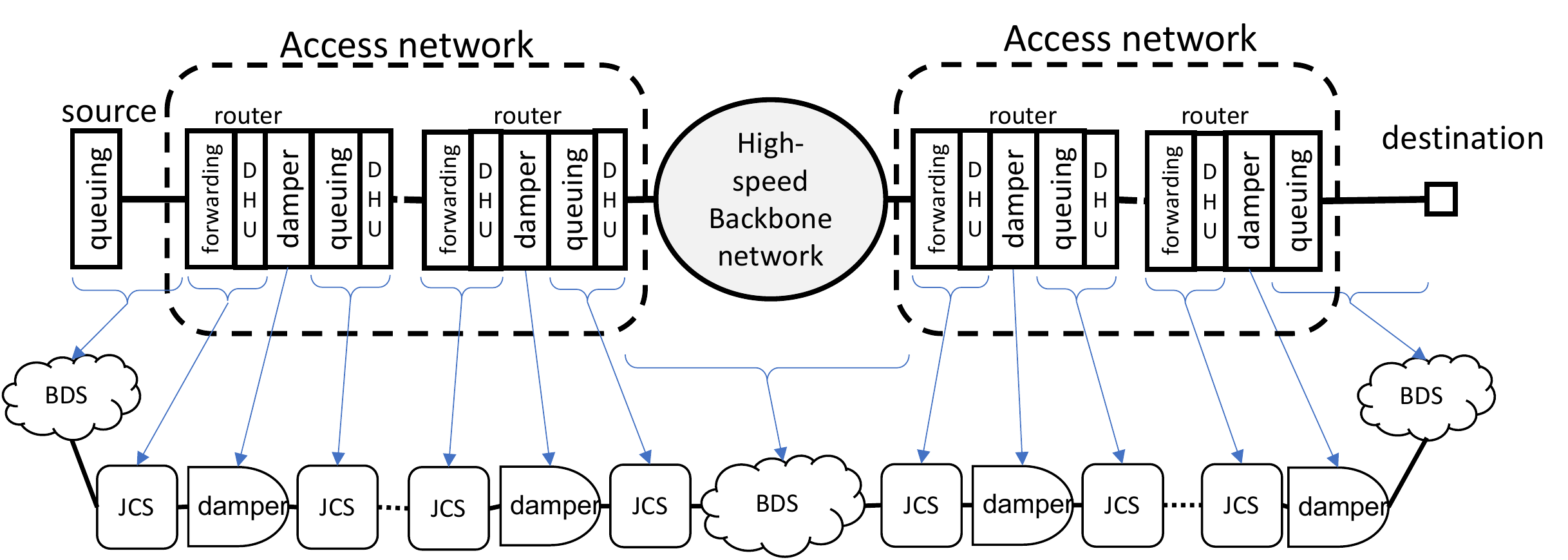}
	\caption{A large-scale deterministic network example with no jitter removal at backbone network. }
	\label{fig:detnet-usecase}
\end{figure}

\textbf{Example 2.} Fig. \ref{fig:detnet-usecase} shows an example flow path within a large-scale deterministic network. Assume that the backbone network has relatively low delay (because of high-speed links, e.g. $100Gbps$ or more, worst-case delays tend to shrink \cite{mathis2019deprecating}) and then the main source of jitter is the access network. For a given class of traffic, we want to remove the jitter imposed by the access network, in particular the forwarding plane and output queuing of each access router (each is treated as a JCS);
therefore, each of these should have a DHU and a damper upstream of the output queuing system. In this example, the backbone network is modelled as a BDS; also, the source is unaware of any downstream damper and does not have a DHU and hence treated as a BDS. The jitter imposed by the access network is removed, but not the jitter caused by the backbone. The different clocks need not be synchronized and the backbone nodes are unmodified.

\begin{figure}[h]
	\centering
	\includegraphics[width= \linewidth]{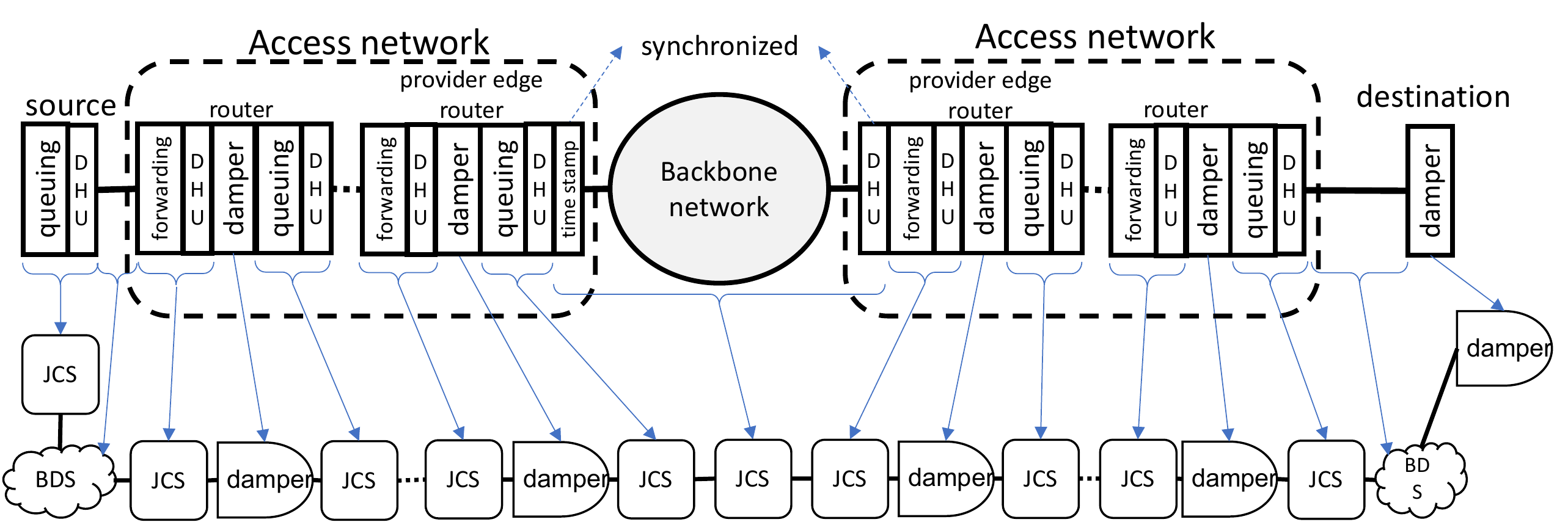}
	\caption{A large-scale deterministic network example with jitter removal at backbone network. }
	\label{fig:detnet-usecase-backbone}
\end{figure}

\textbf{Example 3.}
We continue in Fig. \ref{fig:detnet-usecase-backbone} with the previous example but assume now that, for some class of traffic with very low jitter requirement, the jitter induced by the backbone should be compensated. In such a case, we need to treat the backbone network as a JCS, i.e., we need to time stamp the arrival of each time-sensitive packet to the backbone and modify their damper header at the departure from the backbone. This can be done as in Fig. \ref{fig:detnet-usecase-backbone}, where, at the upstream provider edge (PE) router, a time stamping unit should be added that inserts a field in the packet header equal to the departure time of each time-sensitive packet from the PE router in its local time (this operation can be done within the upstream DHU unit to avoid placement of a time-stamping unit); then at the egress downstream PE router, a DHU is placed that reads the departure time of the packet from packet header, removes it from packet header, computes earliness with its local clock and finally modifies the damper header. In this case, differently from previous examples, the time stamping and DHU are performed with different clocks; therefore, the PE routers should be time-synchronized, as otherwise the computation of earliness is impossible (time synchronization is never absolute and, in sections \ref{sec:clock} and \ref{sec:damper-header}, we analyze how to account for clock non-idealities). The jitter induced by the backbone network is compensated in the damper placed in the downstream PE router and hence removed. The PE routers must be time-synchronized (in provider networks, they typically are); backbone nodes are unmodified but deterministic packets carry an additional header for timestamps.

\subsection{Assumptions on the Clocks}\label{sec:clock}
 We call $\mathcal{H}_{\mathrm{TAI}}$ the perfect clock, i.e. the international atomic time (TAI\footnote{Temps Atomique International}). % We assume that it represents a continuous quantity. 
 In practice, the local clock of a system deviates from the perfect clock \cite{thomas_time_2020}. Typically the JCSs upstream of a damper operate with different clocks than the damper itself, and this can affect the performance of the damper as we see in Section~\ref{sec:analysis-noFIFO}.
In time-sensitive networks, clocks can be synchronized or non-synchronized. Non-synchronized clocks are independently configured and do not interact with each other; this corresponds to the \textit{free-running} mode in \cite[Section 4.4.1]{g810}. When clocks are synchronized, using methods like Network Time Protocol (NTP) \cite{ntp}, Precision Time Protocol (PTP) \cite{ptp}, WhiteRabbit \cite{whiterabbit}, Global Positioning System (GPS) \cite{gps}, the occurrence of an event, when measured with different clocks, is bounded by the \emph{time error bound} ($\sim 1\mu$s or less in PTP, WhiteRabbit, and GPS; $\sim 100$~ms in NTP).

We follow the clock model in~\cite{thomas_time_2020}, which applies to time-sensitive networks. Consider a clock $\mathcal{H}_i$ that is either synchronized with time error bound $\omega$, or not synchronized (in which case we set $\omega=+\infty$). Let $d^{\mathcal{H}_{i}}$ [resp. $d^{\mathcal{H}_{\mathrm{TAI}}}$] be a delay measurement done with clock $\mathcal{H}_i$ [resp. in TAI], then~\cite{thomas_time_2020}:
\begin{align}\label{eq:gamma}
\nonumber d^{\mathcal{H}_{\mathrm{TAI}}}-d^{\mathcal{H}_{i}} &\leq \min\left((\rho-1)d^{\mathcal{H}_{i}}+\eta,2\omega\right),\\
d^{\mathcal{H}_{\mathrm{TAI}}}-d^{\mathcal{H}_{i}} &\geq - \min\left(\left(1-\frac{1}{\rho}\right)d^{\mathcal{H}_{i}}+\frac{\eta}{\rho},2\omega\right),
\end{align}
where $\rho$ is the \emph{stability bound} and $\eta$ the \emph{timing-jitter bound} of the clock $\mathcal{H}_{i}$. Note that this set of bounds is symmetric, i.e. we can exchange the roles of $\mathcal{H}_{i}$ and $\mathcal{H}_{\mathrm{TAI}}$ in \eqref{eq:gamma}. We assume that the parameters $\omega, \rho,\eta$ are valid for all clocks in the network, i.e. we consider network-wide time-error, stability and time-jitter bounds. When a flow has $\alpha^{\mathcal{H}_i}$ as arrival curve with clock $\mathcal{H}_{i}$, then an arrival curve in TAI is~\cite{thomas_time_2020}:
\begin{align}\label{eq:arr-sync}
\alpha^{\mathcal{H}_{\mathrm{TAI}}}:t\rightarrow \alpha^{\mathcal{H}_i}\left(\min\left\{\rho t+\eta,t+2\omega\right\}\right).
\end{align}

In a TSN network, $\rho-1=10^{-4}$ \cite[Annex B.1.1]{802.1AS_ieee_2011} and $\eta=2$ns \cite[Annex B.1.3.1]{802.1AS_ieee_2011}]; if the network is synchronized with gPTP (generic PTP) then $\omega=1\mu$s \cite[Section B.3]{802.1AS_ieee_2011} and if it is not synchronized then $\omega=+\infty$.
\subsection{Delay Jitter}\label{sec:jitter}
For a given flow, call $d_n$ the delay of packet $n$, measured in TAI. The ``worst-case delay" of the flow is $\max_{n}\{d_n\}$ where the max is over all non-lost packets sent by the flow during its lifetime. Similarly, the ``best-case delay" of the flow is $\min_{n}\{d_n\}$. The ``delay jitter" (also called ``jitter") is the difference, i.e., $
V = \max_{n}\{d_n\} - \min_{m}\{d_m\},
$ so that $d_m-d_n\leq V$ for any $m,n$. Delay jitter is called IP Packet Delay Variation in RFC 3393 \cite{rfc3393}.
\section{Taxonomy of Dampers}\label{sec:dm}
As mentioned earlier, designing a damper is a challenge and there exist very different implementations. In this section we classify such implementations in a manner that will be useful for our timing analysis. In the rest of this paper we call ``eligibility time" the time at which a damper releases a packet, as in most implementations the packet is not actually moved, but simply made visible to the next processing element.

\subsection{Dampers without FIFO constraint}\label{sec:nonfifo-damper}
 An \textit{ideal} damper delays a packet by exactly the amount required by the damping header.
 Consider a packet $n$ with damper header $H_n$ that arrives at local time $Q_n$ to a damper. The theoretical eligibility time $\tilde{E}_n$ for the packet is:
\begin{align}\label{eq:elig-ideal-damper}
\tilde{E}_n &= Q_n + H_n,
\end{align} and the ideal damper releases the packet at time $\tilde{E}_n$.
Jitter-control EDF \cite{verma_delay_1991} is an ideal damper, used in combination with an EDF scheduler.

Many other implementations of dampers use some tolerance for the packet release times, due to the difficulty of implementing exact timings. We call damper \textit{with tolerances} $(\Delta^{\textrm{L}},\Delta^{\textrm{U}})$ a damper such that the actual eligibility time $E_n$, of packet $n$, in local time, satisfies:
\begin{align}\label{eq:elig-concrete-damper}
\tilde{E}_n-\Delta^{\textrm{L}} \leq E_n &\leq \tilde{E}_n+\Delta^{\textrm{U}}.
\end{align}
The tolerances can vary from hundreds of nanosecond to a few microsecond based on implementation. Hereafter, we study two instances of dampers with tolerances $(\Delta^{\textrm{L}},\Delta^{\textrm{U}})$, namely, RCSP \cite{zhang_rate-controlled_1993} and RGCQ \cite{rgcq}.

RCSP is an implementation of a damper in relation with static-priority scheduler; each queue of the scheduler is implemented as a linked list and the damper is implemented as a set of linked lists and a calendar queue \cite{brown_calendar_1988}. A calendar queue contains a clock and a calendar where each calendar entry points to an array of linked lists (each for one priority queue). The clock ticks every fix interval $\Delta$. On each clock tick, the linked list that the current clock time of the calendar points is appended to the corresponding priority queue of the scheduler. Whenever a packet $n$ arrives, its theoretical eligibility time is computed based on \eqref{eq:elig-ideal-damper}; then the actual eligibility time of the packet, $E_{n,\mathrm{RCSP}}$, is computed by rounding down the theoretical eligibility time,
%\begin{align}\label{eq:rcsp-actu}
$E_{n,\mathrm{RCSP}} = \Delta\floor{\frac{\tilde{E}_n}{\Delta}}$.
%\end{align}
Then, if $E_{n,\mathrm{RCSP}}$ is equal to the current clock time, it is appended to the corresponding priority queue of the scheduler; otherwise, it is appended to the linked list that the entry $E_{n,\mathrm{RCSP}}$ of the calendar points to. The computation of theoretical eligibility time is done with some errors in acquiring true local-time on packet arrival and in computation due to finite precision arithmetic that is bounded by $\varepsilon$ (typically, in the order of a few nanoseconds). We can see that $E_{\mathrm{RCSP}}$ satisfies \eqref{eq:elig-concrete-damper} when $(\Delta^{\textrm{L}},\Delta^{\textrm{U}})=(\Delta+\varepsilon,\varepsilon)$.

RGCQ, inspired by the idea of Carousel \cite{saeed_carousel_2017}, is an implementation of a damper combined with a queuing system of a scheduler; in other words, each queue of a scheduler is replaced with an RGCQ. An RGCQ consists in a timer and a set of gate-control queues (GCQs). By default, the GCQs are closed and are assigned unique increasing \textit{openTime}s with interspacing of $\Delta$. A GCQ is opened whenever the timer reaches to its openTime and is closed after it is emptied or being opened for a fixed amount of \textit{expiration} time; when a GCQ is closed, its openTime is set as the largest openTime+$\Delta$.
%	 increased by a fixed amount of \textit{cycle} time.
The scheduler selects a packet for transmission from an open GCQ with smallest openTime. Whenever a packet $n$ arrives, its theoretical eligibility time is computed based on \eqref{eq:elig-ideal-damper}; then the actual eligibility time of the packet, $E_{n,\mathrm{RGCQ}}$, is computed by rounding up the theoretical eligibility time, i.e.,
$
%\begin{align}
E_{n,\mathrm{RGCQ}} = \Delta\ceil{\frac{\tilde{E}_n}{\Delta}}$.
%\end{align}
Then, the packet is enqueued to the GCQ whose openTime is $E_{n,\mathrm{RGCQ}}$. Similarly to RCSP, due to timing acquisitions and arithmetic rounding bounded by $\varepsilon$, we see that $E_{\mathrm{RGCQ}}$ satisfies \eqref{eq:elig-concrete-damper} when $(\Delta^{\textrm{L}},\Delta^{\textrm{U}})=(\varepsilon,\Delta+\varepsilon)$.

\subsection{Dampers with FIFO constraint}\label{sec:fifo-damper}
The definition of damper with tolerance given in the previous section does not mention whether the damper preserves packet order, and the satisfaction of \eqref{eq:elig-concrete-damper} does not preclude packet misordering. Indeed, we show in Section~\ref{sec:concrete-reordering} that our two examples of dampers with tolerance, namely RCSP and RGCQ, can cause packet misordering due to clock non-idealities. Such a behavior is not possible with a class of proposed damper designs, which enforce the FIFO constraint, and which we now cover.

\subsubsection{Re-sequencing damper}
We call \emph{re-sequencing damper} with tolerances $(\Delta^{\textrm{L}},\Delta^{\textrm{U}})$ a system that behaves as the concatenation of a damper with same tolerances and a re-sequencing buffer that, if needed, re-orders packets based on the packet order at the entrance of the damper. The packet order is with respect to a flow of interest.

Formally, a system is a re-sequencing damper if there exists a sequence $\bar{E}_n$ such that the release times for the flow of interest, in local time, satisfy:
\begin{align}\label{eq:elig-fifo-damper}
\nonumber \tilde{E}_n-\Delta^{\textrm{L}} &\leq \bar{E}_n \leq \tilde{E}_n+\Delta^{\textrm{U}}, \\
E_1 = \bar{E}_1,~~&E_n = \max\left\{\bar{E}_n, E_{n-1}\right\},
\end{align} where $\tilde{E}_n$ is the theoretical eligibility time defined in \eqref{eq:elig-ideal-damper} and packet numbers $n=1,2,...$ are in order of arrival at the damper.

It follows that such a damper is FIFO for the flow of interest and that\footnote{The converse does not hold, i.e., any system that is FIFO for the flow of interest and satisfies \eqref{eq:elig-fifo-damper-next} is not necessarily a re-sequencing damper.}:
\begin{align}\label{eq:elig-fifo-damper-next}
	\max_{i\leq n}\left\{\tilde{E}_i\right\}-\Delta^{\textrm{L}} \leq E_n &\leq \max_{i\leq n}\left\{\tilde{E}_i\right\}+\Delta^{\textrm{U}}.
\end{align}

We say that a re-sequencing damper is \emph{ideal} if has zero tolerances. Hereafter, we describe two instances of re-sequencing dampers, namely, SCED+ \cite{cruz_sced+:_1998} and FOPLEQ \cite{fopleq}.

SCED+ is an implementation of a damper in combination with SCED scheduling. The damper in \cite{cruz_sced+:_1998} is defined as a conceptual element with tolerance $\Delta$. Accordingly, each packet is assigned an initial eligibility time and a terminal eligibility time where the difference between the two is $\Delta$. In SCED+, the damper ensures that the damper is released after the initial eligibility time and before the terminal eligibility time. In fact, the dampers assigns a tentative eligibility time, $\bar{E}_n$ to a packet $n$, where:
\begin{align}\label{eq:inout-SCED+-single}
\tilde{E}_n-\Delta \leq \bar{E}_n \leq  \tilde{E}_n.
\end{align}
SCED+ assumes that the damper serves packets in FIFO manner; then, the actual eligibility time of the packet $n$ is:
\begin{align}\label{eq:inout-SCED+}
E_1=\bar{E}_1,~~ E_n = \max\left\{\bar{E}_n, E_{n-1}\right\};~~n\geq 2.
\end{align}
so that SCED+ is a re-sequencing damper with tolerances $(\Delta+\varepsilon,\varepsilon)$, where $\varepsilon$ is a bound on the errors on timing acquisition and arithmetic rounding.

FOPLEQ, similarly to RGCQ, is inspired by the architecture of Carousel \cite{saeed_carousel_2017}. Accordingly, it has a set of time-based queues along with a table, called eligibility time table (ETT), for the purpose of preserving the order of packets inside FOPLEQ. Each row in ETT belongs to a flow that has a packet in the Carousel and stores a tentative eligibility time of the latest packet belonging to the corresponding flow. The tentative eligibility time of a packet is obtained by dividing its theoretical eligibility time by $\Delta$ and rounding down the computed value. Consider a packet $n$ of the flow of interest, where number is in the order of arrival at the FOPLEQ. First a theoretical eligibility time is computed using \eqref{eq:elig-ideal-damper}; second, a tentative eligibility time is obtained by rounding down to a multiple of $\Delta$, i.e.
$\bar{E}_n=\Delta\lfloor\frac{\tilde{E}_n}{\Delta}\rfloor$;
then, the actual eligibility time of the packet is the maximum of its tentative eligibility time and the stored tentative eligibility time of the flow of interest in the ETT. The tentative eligibility times correspond to a damper with tolerances $(\Delta+\varepsilon,\varepsilon)$, where $\varepsilon$ is a bound on the errors on timing acquisition and arithmetic rounding, and therefore FOPLEQ is a re-sequencing damper with tolerances $(\Delta+\varepsilon,\varepsilon)$.

\subsubsection{Head-of-line (HoL) damper}
The idea is introduced in \cite{grigorjewMetzgerHossfeldetal_2020}. A HoL damper is implemented as a FIFO queue. When a packet arrives, its arrival time is collected and the packet is stored at the tail of the queue. Only the packet at the head of the queue is examined; if its eligibility time is passed, it is immediately released, otherwise it is delayed and released at its eligibility time. When the head packet is released, it is removed from the damper queue and the next packet (if any) becomes the head of the queue and is examined. When an arriving packet finds an empty queue, it is immediately examined. By construction, packet ordering is preserved.

As before, the model should incorporate some tolerance to account for the timing inaccuracy and for processing times. Unlike with previous damper models, these two things cannot be aggregated because the head-of-line property has the effect that processing times may have an effect over subsequent packets (this is visible in Theorem~\ref{thm:perhop-jcats-delay-fifo}).

Formally, we model a head-of-line damper as follows. It has tolerance parameters $\Delta^{\textrm{L}},\Delta^{\textrm{U}}$ that account for the accuracy of timings, as well as processing bounds $\phi^{\min},\phi^{\max}$ that account for non-zero processing times. We must have $\Delta^{\textrm{L}}\geq0,\Delta^{\textrm{U}}\geq 0$ and $0\leq \phi^{\min}\leq\phi^{\max}$. Packet numbering is with respect to the order of arrivals at the damper and is global for this damper (not per-flow).
We say that a system is a head-of-line damper if the release times $E_n$, in local time, satisfy:
%\begin{eqnarray}
$$\tilde{E}_1-\Delta^{\textrm{L}} \leq E_1 \leq \tilde{E}_1+\Delta^{\textrm{U}},
$$
$$
\max(\tilde{E}_n -\Delta^{\textrm{L}}, E_{n-1})+\phi^{\min} \leq E_n
$$
\begin{equation}
\;\;\;\;\;\;\leq  \max(\tilde{E}_n +\Delta^{\textrm{U}}, E_{n-1})+\phi^{\max},
    \label{eq:elig-hol-damper-1}
\end{equation}
where $\tilde{E}_n$ is the theoretical eligibility time as in \eqref{eq:elig-ideal-damper}.

The definition in \eqref{eq:elig-hol-damper-1} can be explained as follows. First, the eligibility times are obtained with some errors due to timing acquisition and arithmetic rounding. Let $\bar{E}_n$ be the resulting tentative eligibility times, so that
\begin{align}\label{eq:elig-hol-damper-11}
\tilde{E}_n-\Delta^{\textrm{L}} &\leq \bar{E}_n \leq \tilde{E}_n+\Delta^{\textrm{U}}.
\end{align}
Second, packet $n$ is examined only when packet $n-1$ is released, and this action takes a processing time $\phi_n \in [\phi^{\min},\phi^{\max}]$. The actual release time is therefore
\begin{align}\label{eq:elig-hol-damper-12}
E_n & = \max(\bar{E}_n, E_{n-1})+\phi_n.
\end{align}
Using Lemma~\ref{lem:hol-definition} in Appendix~\ref{app:dm-extensive} with $a=E_{n-1}$, $x^{\min}=\phi^{\min}$, $x^{\max}=\phi^{\max}$,  $y^{\min}=\tilde{E}_n-\Delta^{\textrm{L}}$,  $y^{\max}=\tilde{E}_n+\Delta^{\textrm{U}}$, $x=\phi_n$, $y=\bar{E}_n$ and $z=E_n$ we obtain that
 \eqref{eq:elig-hol-damper-11} and \eqref{eq:elig-hol-damper-12} imply \eqref{eq:elig-hol-damper-1}; conversely, if  \eqref{eq:elig-hol-damper-1} holds, there exists sequences $\bar{E}_n$ and $\phi_n\in [\phi^{\min},\phi^{\max}]$ such that \eqref{eq:elig-hol-damper-11} and \eqref{eq:elig-hol-damper-12} hold.

If the tolerances and processing bounds are all equal to $0$, then the HoL damper is called \emph{ideal}. It follows immediately from \eqref{eq:elig-hol-damper-1} and \eqref{eq:elig-fifo-damper} that an ideal HoL damper is the same as an ideal re-sequencing damper.

In \cite{grigorjewMetzgerHossfeldetal_2020}, Jitter-control ATS is presented as an ideal head-of-line damper in combination with Asynchronous Traffic Shaping \cite{ieee8021qcr} within a switch where each FIFO queue is shared among all time-sensitive flows that come from the same input port, have the same class, and go to the same output port. In \cite{grigorjewMetzgerHossfeldetal_2020}, the authors implicitly assume that the tolerances and processing times are zero and therefore ignore them in their analysis. This assumption might not hold in practical cases, specifically when a large number of packets become eligible at the same time in a jitter-control ATS. The effect of non-zero tolerances and processing times appear in Theorem~\ref{thm:perhop-jcats-delay-fifo} and is illustrated numerically in Section~\ref{sec:eval}.

\subsection{Damper Header Computation}\label{sec:damper-header}
In this subsection, we first describe the operation of damper header update unit. Then, we discuss the possible sources of error in the computation.
\subsubsection{DHU unit operation}
The DHU unit of a JCS computes the earliness of a packet and updates the damper header. A classical approach to compute the earliness is to first measure the actual delay of the packet in the JCS with the clock of DHU unit; then set the earliness as the difference between the known delay bound $\delta$ of the system for this class of traffic and the actual delay of the packet \cite{verma_delay_1991,zhang_rate-controlled_1993}. More precisely, for a packet $n$, its arrival time is time stamped with local clock ${\mathcal{H}_{\mathrm{TS}}}$ and stored locally (Examples 1 and 2 in Section \ref{sec:term}) or delivered by the packet (Example 3 in Section \ref{sec:term}); let $\tilde{A}_n^{\mathcal{H}_{\mathrm{TS}}}$ denote the stored/delivered value. Then the DHU unit time stamps the departure time of the packet with its local clock $\mathcal{H}_{\mathrm{DHU}}$; let $\tilde{W}_n^{\mathcal{H}_{\mathrm{DHU}}}$ denote the departure time. Then, the DHU unit computes the earliness of the packet as
\begin{align}\label{eq:earliness}
\mathrm{earliness}_n= \delta - \left(\tilde{W}_n^{\mathcal{H}_{\mathrm{DHU}}}-\tilde{A}_n^{\mathcal{H}_{\mathrm{TS}}}\right).
\end{align}
The last step for the DHU unit is to update the damper header that is equal to the current damper header incremented by the computed earliness, and write the result in the damper header field. Then the packet leaves the JCS. In the case that the JCS is connected to an output link, the departure time of a packet is the complete packet transmission and thus the packet header is accessible to write the damper header just before packet transmission. Therefore, the start of transmission time of the packet is time stamped ($T_n^{\mathcal{H}_{\mathrm{DHU}}}$) and the transmission time is inferred as $\tilde{\tau}_n^{\mathcal{H}_{\mathrm{DHU}}} = \left\{\frac{l_n}{c}\right\}^{\mathcal{H}_{\mathrm{DHU}}}$ with $l_n$ as the packet length and $c$ as the transmission rate. Then, we set the departure time to $\tilde{W}_n^{\mathcal{H}_{\mathrm{DHU}}}=T_n^{\mathcal{H}_{\mathrm{DHU}}}+\tilde{\tau}_n^{\mathcal{H}_{\mathrm{DHU}}}$ and compute the earliness using \eqref{eq:earliness}. This method of damper header computation is used in most of the existing damper variants like RCSP \cite{zhang_rate-controlled_1993}, FOPLEQ \cite{fopleq} and jitter-control ATS \cite{grigorjewMetzgerHossfeldetal_2020}. We call this the \emph{default} method of damper header computation.

Recently, \cite{rgcq} proposed a subtle change in the computation of earliness when a JCS comes immediately after a damper with tolerances $(\Delta^{\mathrm{L}},\Delta^{\mathrm{U}})$. In particular, they suggest to time stamp the theoretical eligibility time $\tilde{E}_n$ of packet from the damper instead of the arrival time to the JCS; as a consequence, the jitter imposed by the tolerance of the damper is compensated by the next upstream damper. In such proposal, note that the delay upper-bound between the theoretical eligibility time to the arrival time to the JCS (i.e., the actual eligibility time from the damper) should be added to the earliness; by \eqref{eq:elig-concrete-damper}, this upper bound is $\Delta^{\mathrm{U}}$. Hence, the earliness for theoretical eligibility time stamping is:
\begin{align}\label{eq:earliness-1}
\mathrm{earliness}_{n}= \delta + \Delta^{\mathrm{U}}- \left(\tilde{W}_n^{\mathcal{H}_{\mathrm{DHU}}}-\tilde{A}_n^{\mathcal{H}_{\mathrm{TS}}}\right).
\end{align}
We call this method of damper header computation \emph{TE time-stamping}.

\subsubsection{Errors in damper header computation}
\label{sec:errors}

The DHU unit computes a damper header equal to the current damper header incremented by the computed earliness, and write the result in the damper header field. This step is imperfect due to finite precision arithmetic and finite resolution of the damper header. The corresponding error is
$
e_{\mathrm{update},n} = H_n - \tilde{H}_n,
$
where $\tilde{H}_n$ is the theoretical value of the damper header and $H_n$ is the actual value written in the packet.

In the computation of earliness (\eqref{eq:earliness} and \eqref{eq:earliness-1}), when $\tilde{A}_n^{\mathcal{H}_{\mathrm{TS}}}$ is delivered within a packet header field (Example 3 in Section \ref{sec:term}), there is some error induced due to the finite resolution of the header field. The corresponding error is
$
e_{\mathrm{ts},n} = {A'}_n^{\mathcal{H}_{\mathrm{TS}}} - \tilde{A}_n^{\mathcal{H}_{\mathrm{TS}}},
$
where ${A'}_n^{\mathcal{H}_{\mathrm{TS}}}$ is the time stamped value at the packet arrival to the JCS.

As discussed earlier, when the JCS is connected to a transmission link, the DHU unit can infer the transmission time by dividing the packet length over the nominal transmission rate of the link. Due to transmission of preamble and inexact knowledge of actual transmission rate, the inference of transmission time is done with some error
$
e_{\mathrm{tran},n} = \tilde{\tau}_n^{\mathcal{H}_{\mathrm{DHU}}} - \tau_n^{\mathcal{H}_{\mathrm{DHU}}},
$
where $\tau_n^{\mathcal{H}_{\mathrm{DHU}}}$ and $\tilde{\tau}_n^{\mathcal{H}_{\mathrm{DHU}}}$ are the actual and inferred transmission times. The error $e_{\mathrm{tran},n}$ can go up to tens of nanoseconds \cite{intelMegaCore,10gMac}.

Acquiring the true local-time on packet arrival and within the DHU unit usually comes with an error. We define the clock acquisition error as:
$
e_{\mathrm{acq},n} = \left({W}^{\mathcal{H}_{\mathrm{DHU}}}_n - \tilde{W}^{\mathcal{H}_{\mathrm{DHU}}}_n\right)+\left({A}^{\mathcal{H}_{\mathrm{TS}}}_n - {A'}^{\mathcal{H}_{\mathrm{TS}}}_n\right),
$
where $A^{\mathcal{H}_{\mathrm{TS}}}_n$ and $W^{\mathcal{H}_{\mathrm{DHU}}}_n$ are the true local times on packet arrival and departures.

The two clocks $\mathcal{H}_{\mathrm{DHU}}$ and $\mathcal{H}_{\mathrm{TS}}$ are often the same (Examples~1 and~2 in Section \ref{sec:term}), but not always (Example~3 in Section \ref{sec:term}). We select $\mathcal{H}_{\mathrm{DHU}}$ as the reference clock of a JCS to compute damper header; then we define the error with respect to the reference clock as:
$
e_{\mathrm{clk},n} = A^{\mathcal{H}_{\mathrm{DHU}}}_n - A^{\mathcal{H}_{\mathrm{TS}}}_n,
$
where $A^{\mathcal{H}_{\mathrm{DHU}}}_n$ is the time that would be displayed at packet arrival if $\mathcal{H}_{\mathrm{DHU}}$ would be used. If the clocks are the same, $e_{\mathrm{clk}}=0$; if the clocks are synchronized with error $\omega$ with respect to TAI, $|e_{\mathrm{clk}}|\leq 2\omega$; and finally if the clocks are not synchronized, $e_{\mathrm{clk}}$ can get arbitrary large, which is incompatible with the goal of removing jitter. Therefore, in this paper, we assume that both clocks $\mathcal{H}_{\mathrm{TS}}$ and $\mathcal{H}_{\mathrm{DHU}}$ are either one and the same, or are synchronized.

To summarize, the value of a damper header, as written in a packet $n$, suffers from some error $e_n$ equal to:
%\begin{align}\label{eq:error}
$
e_n = e_{\mathrm{update},n}+e_{\mathrm{ts},n}+e_{\mathrm{tran},n} +e_{\mathrm{acq},n}+e_{\mathrm{clk},n}
$.
%\end{align}
Each of these sources of error can be bounded, depending on the technology used by the routers and switches. The variable $\epsilon$ denotes an upper bound on the error $e_n$, i.e., $|e_n| \leq \epsilon$. When $\mathcal{H}_{\mathrm{DHU}}$ and $\mathcal{H}_{\mathrm{TS}}$ are the same, $\epsilon$ is typically of the order of tens of nanoseconds; if they are synchronized, $\epsilon$ is mainly dominated by the time error bound (e.g. $\epsilon= 2~\mu$s for gPTP).

\section{Delay Analysis of Dampers without FIFO constraints} \label{sec:analysis-noFIFO}
\begin{figure}[th]
	\centering
	\includegraphics[width=0.9 \linewidth]{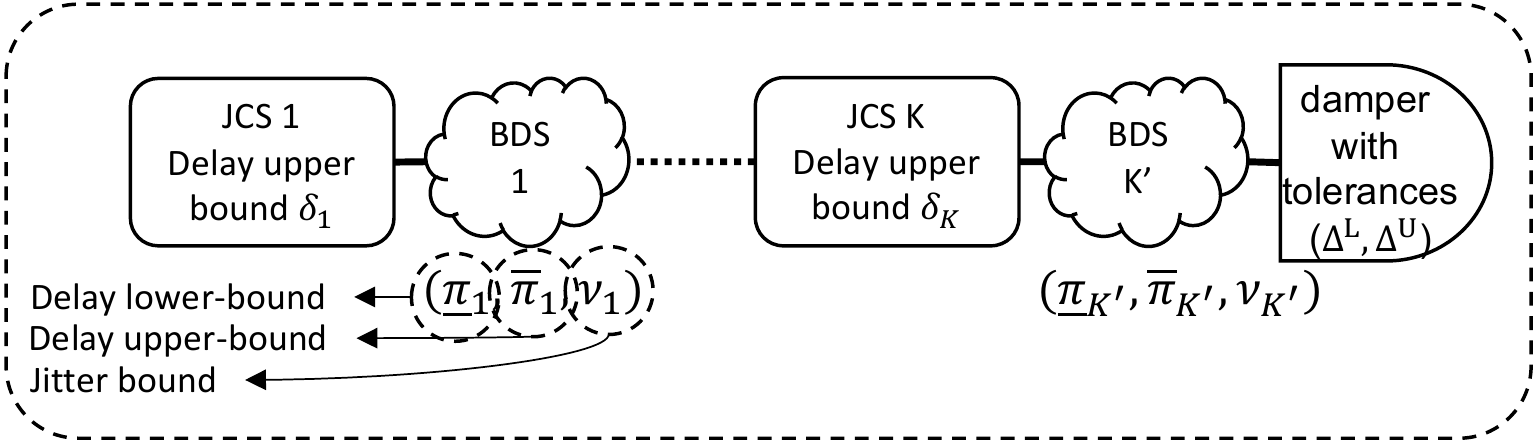}
	\caption{Structure of a block whose delay and jitter bounds are computed in Theorem \ref{thm:perhop-concrete-delay}.}
	\label{fig:hop-concrete}
\end{figure}

In this section we study the end-to-end delay and jitter of a flow when dampers without FIFO constraint are used. The first step is
to decompose a flow path into a set of blocks that can be analyzed separately. Every block is as in Fig.~\ref{fig:hop-concrete}; it contains a number of JCSs and BDSs and ends in a damper with tolerance. 
The second step, given in the rest of this section, is to give delay and jitter bounds for a flow through such a block.
The bounds are valid whether the JCSs and the BDSs of the block are FIFO or not. The last step, to obtain  end-to-end results, simply consists in summing up the delays and jitters of every block and, possibly, of remaining BDSs. For example, in Fig.~\ref{fig:detnet-usecase}, from source to the first router and from the queuing system of the last router to the destination are BDSs and the rest are decomposed in blocks as in Fig.~\ref{fig:hop-concrete}.

In the following we give delay and jitter bounds for a block as in Fig.~\ref{fig:hop-concrete}. Theorem \ref{thm:perhop-concrete-delay} gives the result for dampers without FIFO constraint when the default mode of header computation is used (as explained in Section~\ref{sec:damper-header}) and Theorem~\ref{thm:e2e-concrete} when TE time-stamping is used. In both cases, we capture the effect of errors  and non-ideal clocks. We also illustrate cases where the errors and non-ideal clocks make a major contribution to the jitter bound.

\begin{theorem}\label{thm:perhop-concrete-delay}
	Consider a flow of interest that traverses the block in Fig. \ref{fig:hop-concrete}. The block contains a sequence of JCSs and BDSs and terminates in a damper with tolerances ($\Delta^{\mathrm{L}},\Delta^{\mathrm{U}}$). Assume that the clock of every system has stability bound $\rho$, timing-jitter bound $\eta$ and time-error bound $\omega$ with respect to TAI (Section~\ref{sec:clock}). Assume that JCS $i$ has a delay upper bound $\delta_i$, which is used for damper header computation, in its local time. Also, assume that the BDS $j$ has delay lower and upper bounds $\underline{\pi}_j^{\mathcal{H}_{\mathrm{TAI}}}$ and $\overline{\pi}_j^{\mathcal{H}_{\mathrm{TAI}}}$ and jitter bound $\nu_j^{\mathcal{H}_{\mathrm{TAI}}}$, all in TAI. Then, the delay of a packet from entrance to the exit of the block, in TAI, is upper-bounded by $\overline{D}$, low-bounded by $\underline{D}$ and has jitter bound $V$, with
	\begin{align}
	\nonumber\overline{D}&=\sum_{j=1}^{K}\delta_j+\sum_{j=1}^{K'} \overline{\pi}_j^{\mathcal{H}_{\mathrm{TAI}}}+\Delta^{\mathrm{U}}+K\epsilon +\overline{\psi},\\
	\nonumber \underline{D}&=\sum_{j=1}^{K}\delta_j+\sum_{j=1}^{K'}\underline{\pi}_j^{\mathcal{H}_{\mathrm{TAI}}} -\Delta^{\mathrm{L}}-K\epsilon -\underline{\psi},\\
	\nonumber V&=\sum_{j=1}^{K'}\nu_j^{\mathcal{H}_{\mathrm{TAI}}}  + \Delta^{\mathrm{U}}+\Delta^{\mathrm{L}} + 2K\epsilon+\overline{\psi}+\underline{\psi},
	\end{align}
where $\overline{\psi}$ and $\underline{\psi}$ are due to clock non-idealities,
\begin{align}\label{eq:psi}
\nonumber \overline{\psi}&=\min\Big((\rho-1)\big(\Delta^{\mathrm{U}}+\sum_{j=1}^{K}(\delta_j+\epsilon)\big)+ (K+1)\eta\\
\nonumber&,2(K+1) \omega\Big),\\
\nonumber\underline{\psi} &= \min\Big(\big(1-\frac{1}{\rho}\big)\big(-\Delta^{\mathrm{L}}+\sum_{j=1}^{K}\left(\delta_j-\epsilon\right)\big)+\frac{(K+1)\eta}{\rho}\\
&,2(K+1) \omega\Big).
\end{align}

The bounds are tight, i.e., for any tolerances ($\Delta^{\mathrm{L}},\Delta^{\mathrm{U}}$), every $\delta_i$, any $\underline{\pi}_j^{\mathcal{H}_{\mathrm{TAI}}}$, $\overline{\pi}_j^{\mathcal{H}_{\mathrm{TAI}}}$, $\nu_j^{\mathcal{H}_{\mathrm{TAI}}}$, there is a system and two individual execution traces such that in one of them a packet experiences a delay of $\overline{D}$ and in the other one  a packet experiences a delay of $\underline{D}$.
\end{theorem}

The proof is in Appendix \ref{proof:perhop-concrete-delay}.
\begin{remark}
	The second arguments of the $\min(,)$ functions in \eqref{eq:psi} capture the impact of clock time error bounds when all the $K+1$ clocks ($K$ JCSs and one damper with tolerance) are different from each other. If some systems share a common clock, so that there are $X\leq K$ different clocks in total, the second argument of the $\min(,)$ functions should be replaced by $2(X+1)\omega$.
\end{remark}

Hereafter, we provide an application of Theorem \ref{thm:perhop-concrete-delay} to obtain delay and jitter bounds for the three examples of Section \ref{sec:term}. Then we compare the bounds with the \textit{basic} bounds obtained when assuming that clocks are perfect, i.e. by summing the tolerances of dampers and the jitters of BDSs.

\textbf{Example 1.}	Consider Example 1 in Section \ref{sec:term} for a flow that traverses $6$ switches. Assume that the switching fabrics (as JCSs) have a delay upper-bound of $2~\mu$s and the delay bound at each queuing system (as a JCS) is $250~\mu$s. Suppose that the error $\epsilon=50$~ns and all the dampers are RCSP with $(\Delta^{\mathrm{L}},\Delta^{\mathrm{U}})=(1~\mu$s$, 2$~ns$)$. Assume the propagation delay (as a BDS) is fixed and equal to $5~\mu$s for all links. Assume first that the clocks of the switches are not synchronized, i.e. we set the time-error bound $\omega$ to $\infty$ in \eqref{eq:psi}. Then, by applying Theorem \ref{thm:perhop-concrete-delay} from source to the output of the first damper, the delay upper-bound is $\overline{D}=257.13~\mu$s; the delay jitter is $V=1.262~\mu$s of which $200$~ns is due to errors and $62$~ns is due to non-ideal clocks. The basic jitter bound is $1.002~\mu$s and is due to the tolerance of the first damper. We can see that the error and non-ideal clocks add $262$~ns ($26\%$) to the basic jitter bound. The end-to-end delay and jitter bounds are computed by summing up the delay and jitter bounds from the output of one damper to the output of the next downstream damper until the destination; this gives an end-to-end delay upper-bound  of $1.799$~ms and end-to-end jitter bound of $8.834~\mu$s. The values remain the same if we assume next that the clocks of the switches are synchronized with a time-error bound of $1~\mu$s, as is typical in IEEE TSN systems.

\textbf{Example 2.}
	Consider Example 2 in Section \ref{sec:term} for a flow that traverses four access routers to reach the backbone and traverses four other access routers to reach to the destination. Assume that 1) the queuing delay at source has a delay upper-bound of $100~\mu$s and the  jitter bound of $80~\mu$s, 2) the delay upper-bound at the output queuing and packet forwarding of each access router are $500~\mu$s and $5~\mu$s and the output queuing has jitter of $100~\mu$s, 3) the backbone network has a delay upper-bound of $30$~ms and jitter bound of $1$~ms, 4) the propagation delay is $10~\mu$s, 5) the error $\epsilon=50$~ns and 6) all dampers are RCSP with $(\Delta^{\mathrm{L}},\Delta^{\mathrm{U}})=(1~\mu$s$, 2$~ns$)$. Then, by using Theorem \ref{thm:perhop-concrete-delay}, the end-to-end delay upper-bound is $\overline{D}_{\mathrm{e2e}}=34.211$~ms; delay jitter is $V_{\mathrm{e2e}}=1.190$~ms of which $1.5~\mu$s is due to the errors and $800$~ns is due to non-ideal clocks. The basic jitter bound is $1.188$~ms that is due to the tolerance of the dampers, as well as the BDSs, i.e., the backbone network, the source output queuing and the output queuing of the last access router (before the destination).	We can see that here the effect of the errors and non-ideal clocks is negligible due to the jitter of the BDSs captured by the basic jitter bound.

\textbf{Example 3.}
	Consider Example 3 in Section \ref{sec:term} for a flow that traverses four access routers to reach the backbone and traverses four other access routers to reach to the destination. Also, consider the same numerical assumptions as the previous example. We want to remove  jitter of the backbone network using time stamping at the upstream PE router. Assume that the PE routers are synchronized with error bound of $1~\mu$s; hence the error damper header computation at the downstream PE router of the backbone network is bounded by $2.05~\mu$s (the error at the other JCSs is bounded by $50$~ns).	Then, by using Theorem \ref{thm:perhop-concrete-delay}, the end-to-end delay upper-bound is $\overline{D}_{\mathrm{e2e}}=34.216$~ms; delay jitter is $V_{\mathrm{e2e}}=21.74~\mu$s of which $5.8~\mu$s is due to the errors and $6.92~\mu$s is due to non-ideal clocks. Comparing to the previous example, the basic jitter bound is reduced to $9.02~\mu$s as the jitter of the backbone network, queuing at source and the queuing at the last access router are removed. We can see that the errors and clock non-idealities add $12.72~\mu$s ($141\%$) to the basic jitter bound.

We see from these examples that, when the remaining end-to-end delay-jitter is still large after applying dampers (ms or more, Example 2) then the timing errors and clock non-idealities do not play a significant role and can be ignored. In contrast, for very small residual delay-jitter (Examples 1 and 2, $10~\mu$s or less), ignoring timing errors and clock non-idealities can lead to significant under-estimation.

\begin{remark}
In Example 1, we see that the delay-jitter bound is not affected by the time-error bound, i.e., here, time synchronization does not improve the performance of dampers. We can easily analyze when this is the case, by comparing the terms in the $\min(.)$ functions in \eqref{eq:psi}. We find that time synchronization does not improve the performance of dampers if and only if
\begin{align}
		\label{eq:sync-nonsync} \sum_{j=1}^{K}\delta_j \leq \frac{K+1}{\rho-1}\left(2\omega-\eta\right)-\Delta^{\mathrm{U}}-K\epsilon.
	\end{align}

\noindent It follows that if $\sum_{j=1}^{K}\delta_j > \frac{2}{\rho-1}\left(2\omega-\eta\right)$, the time error bound affects the delay and jitter bounds of Theorem \ref{thm:perhop-concrete-delay}; i.e., it is the relation between the delay (not delay-jitter) bound and the time-error bound that matters (see Table~\ref{table:sync-nonsync}).

\begin{table}[!t]
	\centering
	\caption{Minimum values of the sum of delay bounds for the JCSs within a block such that clock synchronization improves the delay and jitter bounds in Theorem \ref{thm:perhop-concrete-delay}.}
	\begin{tabular}{|c|c|c|}
		\hline&&\\[-1ex]
		\shortstack{Synchronization \\method}&\shortstack{Time-error bound \\($\omega$)}& \shortstack{Minimum value of\\ $\sum_{j=1}^{K}\delta_j$} \\ [0.5ex]
		\hline &&\\[-0.75ex]
		White Rabbit & $100$ns &$3.96$ms\\ [1ex]
		\hline &&\\[-0.75ex]
		gPTP & $1\mu$s &$39.96$ms\\ [1ex]
		\hline &&\\[-0.75ex]
		NTP & $100$ms &$3.99$s\\ [1ex]
		\hline
	\end{tabular}
	\label{table:sync-nonsync}
\end{table}

\noindent TSN networks are typically synchronized with gPTP ($\omega=1~\mu$s) \cite[Section B.3]{802.1AS_ieee_2011}. Three main delay sensitive classes are Control-Data Traffic (CDT), class A for audio traffic and class B for video traffic. According to TSN documents \cite{ieeeAVB,cdt_delay,tsn-profile-service-provider}, the end-to-end delay requirement for CDT, classes A and B are respectively $100~\mu$s in $5$ hops, $2$~ms and $50$~ms in $7$ hops. According to Table \ref{table:sync-nonsync}, the gPTP synchronization does not impact the obtained delay bound using Theorem \ref{thm:e2e-concrete} for CDT and class A. For class B, if we consider that for each block the sum of JCS delay bounds is less than $39.96$~ms, similarly the gPTP synchronization does not play a role. This implies when all switches and the destinations in a TSN network implement dampers with tolerances and the source performs time stamping (as a JCS), then without gPTP synchronization the same performance is achieved.
\end{remark}

In order to provide delay and delay-jitter guarantees to time-sensitive flows, it is often required to bound the burstiness of flows inside the network, which is typically larger than at the source. Finding such bounds may be difficult, and worst-case bounds may be large when there are cyclic dependencies \cite{thomas_on-cyclic_2019}. Here, dampers can help a lot, as shown by the following Corollary, which comes by direct application of the jitter bound in Theorem \ref{thm:perhop-concrete-delay} and \cite[Lemma 1]{mohammadpour2020packet}.

\begin{corollary}\label{col:concrete-ac-prop}
	Consider Fig. \ref{fig:hop-concrete}. Suppose that a flow has $\alpha^{\mathcal{H}_{\mathrm{TAI}}}$ as arrival curve at the entrance of the block. Then an arrival curve at the output of the block, in TAI, is given by
$\alpha_{h}^{\mathcal{H}_{\mathrm{TAI}}}(t) = \alpha^{\mathcal{H}_{\mathrm{TAI}}}\left(t+V\right)$
where $V$ is the jitter bound of the block defined in Theorem~\ref{thm:perhop-concrete-delay}.
\end{corollary}

\begin{exmp}
	In Example 1 of Section \ref{sec:term}, suppose that the flow has leaky bucket arrival curve with rate $16$~Mbps, in TAI, and burstiness $10$~KBytes at source. We computed that the jitter bound is $1.262~\mu$s from source the output of the first damper. Then the arrival curve at the output of the first damper has the same rate and the burstiness is increased by $3$~Bytes. Without damper, the  burstiness increase would be $515$~Bytes: we see that the burstiness increase due to multiplexing is almost entirely removed.
\end{exmp}

\begin{remark}
	The arrival curve constraint at source may be available in its local time rather than in TAI. Then, we can apply \eqref{eq:arr-sync}, to obtain an arrival curve in TAI and then use the result of Corollary \ref{col:concrete-ac-prop}.
\end{remark}

When dampers use TE time-stamping for damper header computation rather than the default method, the delay and jitter bounds computation with TE time-stamping are slightly different than in Theorem~\ref{thm:perhop-concrete-delay}. A JCS is affected only when the upstream damper uses TE time-stamping (otherwise, the bounds are the same). The next theorem gives end-to-end delay and jitter bounds when DHU unit uses TE time-stamping.
%\vspace{-0.4cm}
\begin{figure}[h]
	\centering
	\includegraphics[width= \linewidth]{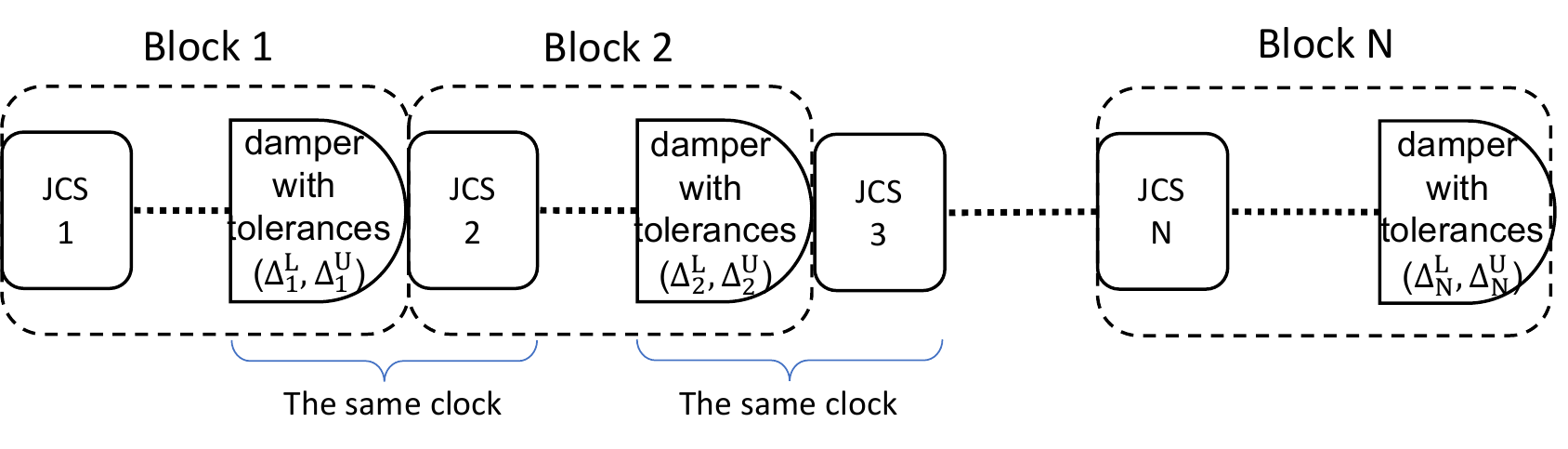}
	\caption{The notations used in Theorem \ref{thm:e2e-concrete}.}
	\label{fig:e2e-concrete}
\end{figure}

%\vspace{-0.3cm}
\begin{theorem}\label{thm:e2e-concrete}
	Consider Fig. \ref{fig:e2e-concrete} where a sequence of $N$ blocks are concatenated and TE time-stamping is used for damper header computation. Assume that clocks follow the description in Section \ref{sec:clock} and damper with tolerances at block $i$ ($i=1,\dots,N-1$) and JCS $i+1$ operate with the same clock. Assume that there are $M$ JCSs in total. Let us denote the the sum of the delay bounds of the JCSs as $\delta_{\mathrm{e2e}}$ and the sum of delay lower and upper bounds and jitter bound of the BDSs as $\underline{\pi}_{\mathrm{e2e}},\overline{\pi}_{\mathrm{e2e}}$ and $\nu_{{\mathrm{e2e}}}$ respectively. Then,
	\begin{align}
	\nonumber\overline{D}_{\mathrm{TE}} &=\delta_{\mathrm{e2e}}+\overline{\pi}_{\mathrm{e2e}}+\sum_{j=1}^{N}\Delta_j^{\mathrm{U}}+M\epsilon+\overline{\Psi}_{\mathrm{TE}},\\
	\nonumber\underline{D}_{\mathrm{TE}} &=\delta_{\mathrm{e2e}}+\underline{\pi}_{\mathrm{e2e}}+\sum_{j=1}^{N-1}\Delta_j^{\mathrm{U}}-\Delta_N^{\mathrm{L}}-M\epsilon-\underline{\Psi}_{\mathrm{TE}},\\
	V_{\mathrm{TE}} &= \nu_{{\mathrm{e2e}}}+\Delta_N^{\mathrm{U}}+\Delta_N^{\mathrm{L}}+2M\epsilon+\overline{\Psi}_{\mathrm{TE}}+\underline{\Psi}_{\mathrm{TE}},
	\end{align}
	where $\overline{\Psi}_{\mathrm{TE}}$ and $\underline{\Psi}_{\mathrm{TE}}$ are the errors due to non-ideal clocks:
	\begin{align}
	\nonumber\overline{\Psi}_{\mathrm{TE}} &= \min\Big((\rho-1)\big(\delta_{\mathrm{e2e}}+\sum_{i=1}^{N}\Delta_i^{\mathrm{U}}+M\epsilon\big)+ (M+N)\eta\\
	\nonumber&,2(M+N) \omega\Big),\\
	\nonumber\underline{\Psi}_{\mathrm{TE}} &=\min\Big(\big(1-\frac{1}{\rho}\big)\big(\delta_{\mathrm{e2e}}-\Delta_N^{\mathrm{L}}+\sum_{i=1}^{N-1}\Delta_i^{\mathrm{U}}+M\epsilon\big)\\
	&+\frac{(M+N)\eta}{\rho},2(M+N) \omega\Big).
	\end{align}
\end{theorem}
The proof is available in Appendix \ref{proof:e2e-concrete}.
\begin{exmp}
Let us redo the end-to-end delay and jitter bounds computation for the three examples of Section \ref{sec:term} using Theorem \ref{thm:e2e-concrete} and compare them with the ones obtained with Theorem \ref{thm:perhop-concrete-delay}. Let us consider the same assumptions made when applying Theorem \ref{thm:perhop-concrete-delay}. We can see that the delay upper-bounds obtained by Theorem \ref{thm:e2e-concrete} are the same as the ones computed after Theorem \ref{thm:perhop-concrete-delay}; however, the end-to-end jitter is reduced. Using Theorem \ref{thm:e2e-concrete}, the end-to-end jitter bound for Example 1 is $V_{\mathrm{Ex1}}=2.834~\mu$s, Example 2 is $V_{\mathrm{Ex2}}=1.183$~ms, Example 3 is $V_{\mathrm{Ex3}}=13.74~\mu$s. The reason for jitter bound reduction by Theorem \ref{thm:e2e-concrete} is the elimination of the jitter imposed by the tolerance of all the intermediate dampers by the next downstream dampers. In examples 1 and 3, the jitter bounds are considerably reduced, by  $68\%$ and $36\%$; however, this is not the case for Example 2 as the main sources of jitter are the BDSs. Furthermore, the jitter imposed by the errors and the non-ideal clocks incorporate $65\%$ and $92\%$ of the the end-to-end jitter bounds computed for examples 1 and 3, which are respectively $2.8$ and $13.74$ times the basic jitter bounds.
\end{exmp}

\section{Packet reordering in Dampers Without FIFO Constraints}
\label{sec:concrete-reordering}
In this Section we show that dampers without FIFO constraint can cause packet misordering, and we quantify the corresponding reordering metrics.

Obviously, a damper modifies packet order if the sequence of theoretical eligibility times is not monotonic. Since the theoretical eligibility time is equal to the arrival time at the JCS plus a constant, this may occur only if the packet order at the entrance to the damper is not the same as at the entrance to the JCS, i.e. this requires the JCS to be non FIFO. But, as we show next, this may occur even if the JCS is FIFO, due to timing inaccuracies.

RGCQ and RCSP are two instances of dampers with tolerance; by design, they avoid packet reordering due to the tolerances by enforcing FIFO behavior after computation of theoretical eligibility times. However, as we show next,  packet reordering may still occur within RGCQ and RCSP due to the errors of damper header computation and non-ideal clocks.

\begin{figure}[h]
	\centering
	\includegraphics[width= 0.5\linewidth]{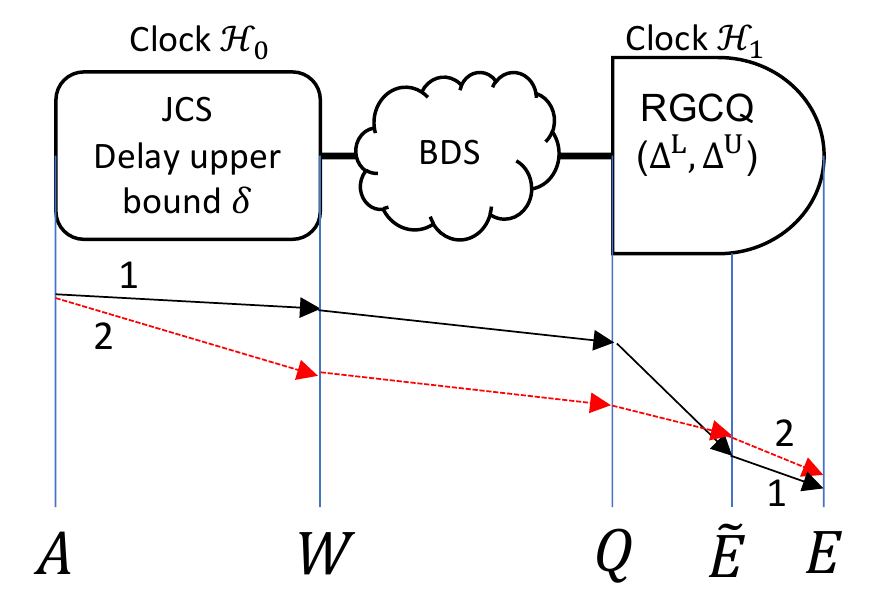}
	\caption{Packet reordering scenario when two packets enter back-to-back to a JCS.}
	\label{fig:reordering-1}
\end{figure}

\textbf{Re-ordering example with RGCQ.}
Consider Fig. \ref{fig:reordering-1} where the damper is RGCQ with tolerances ($\Delta^{\mathrm{L}},\Delta^{\mathrm{U}}$) and clocks are not synchronized. Assume that the JCS represents a FIFO queue connected to a transmission line with a fixed rate and the BDS has zero jitter and represents constant propagation delay (similar to the first hop in Example 1 of Section \ref{sec:term}). Suppose that two packets $1$ and $2$ enter the JCS at the same time while $1$ is prior to $2$. Then packet $1$ leaves before packet $2$. The damper headers in the packets are:
\begin{align}\label{eq:reordering-scen1-header}
	\nonumber H_1 &= \delta -\tilde{\tau}_1^{\mathcal{H}_0},\\
	H_2&=\delta -\tilde{\tau}_1^{\mathcal{H}_0} -\tilde{\tau}_2^{\mathcal{H}_0}=H_1-\tilde{\tau}_2^{\mathcal{H}_0},
\end{align}
where $\tilde{\tau}_i^{\mathcal{H}_0}$ is the inferred transmission times of packets $i \in\{1,2\}$ measured with $\mathcal{H}_0$. Then, the interspacing of the two packet at the output of the JCS is:
\begin{align}
W^{\mathcal{H}_0}_{2}-W^{\mathcal{H}_0}_{1}=\tau_2^{\mathcal{H}_0},
\end{align}
where $\tau_2^{\mathcal{H}_0}$ is the actual transmission time of packet $2$ and $W^{\mathcal{H}_0}_{i}$ is the departure time of packet $i \in\{1,2\}$ from the JCS. Both packets experience the same delay in the BDS. Therefore, the interspacing between the two packets at the entrance of RGCQ when seen with clock $\mathcal{H}_1$ is:
\begin{align}
Q^{\mathcal{H}_1}_{2}-Q^{\mathcal{H}_1}_{1}&=\tau_2^{\mathcal{H}_1} = \tau_2^{\mathcal{H}_0} + \left(\tau_2^{\mathcal{H}_1}-\tau_2^{\mathcal{H}_0}\right),
\end{align}
where $Q^{\mathcal{H}_1}_{i}$ is the arrival time of packet $i \in\{1,2\}$ to RGCQ.
 Then, by \eqref{eq:reordering-scen1-header} and \eqref{eq:elig-ideal-damper}, the difference between the theoretical eligibility times of packets $2$ and $1$ is:
 \begin{align}
 \nonumber \tilde{E}^{\mathcal{H}_1}_{2}- \tilde{E}^{\mathcal{H}_1}_{1} &=  Q^{\mathcal{H}_1}_{2}-Q^{\mathcal{H}_1}_{1} +H_2-H_1 \\
 &= \left(\tau_2^{\mathcal{H}_0} -\tilde{\tau}_2^{\mathcal{H}_0}\right)+ \left(\tau_2^{\mathcal{H}_1}-\tau_2^{\mathcal{H}_0}\right).
 \end{align}
 The difference between the theoretical eligibility times is the sum of the error between actual and inferred transmission time and the measurement difference of packet $2$ transmission time seen from clocks $\mathcal{H}_1$ and $\mathcal{H}_0$. Therefore, if it happens that clock $\mathcal{H}_1$ is faster than $\mathcal{H}_0$ during the transmission time of packet $2$ from the JCS, then $\tau_2^{\mathcal{H}_1}<\tau_2^{\mathcal{H}_0}$, % and assuming uniform error;
 hence
 $%\begin{align}
\tilde{E}^{\mathcal{H}_1}_{2}- \tilde{E}^{\mathcal{H}_1}_{1} < 0,
 $ %\end{align}
 i.e. packet $2$ has smaller theoretical eligible time than packet~$1$. Then, by implementation of RGCQ discussed in Section \ref{sec:dm}, packet~$2$ leaves RGCQ before packet~$1$.
 \begin{remark}
 	In this scenario, reordering occurs because of the difference of speed between the two clocks $\mathcal{H}_0$ and $\mathcal{H}_1$ at the microscopic scale and the error in inferring the transmission time. The earliness of a packet written in the header is measured using the local clock $\mathcal{H}_{0}$ while the delay imposed to the packet is measured in the RGCQ using the local clock $\mathcal{H}_{1}$. Even if both systems are time synchronized, there still remains a small difference in the time measurements performed by the two clocks. Over the transmission time of a packet, there is equal chance that one clocks ticks slightly faster than the other, i.e. there is $50$\% chance that the change of order described in this scenario occurs. 
 \end{remark}

\begin{figure}[h]
	\centering
	\includegraphics[width= 0.8\linewidth]{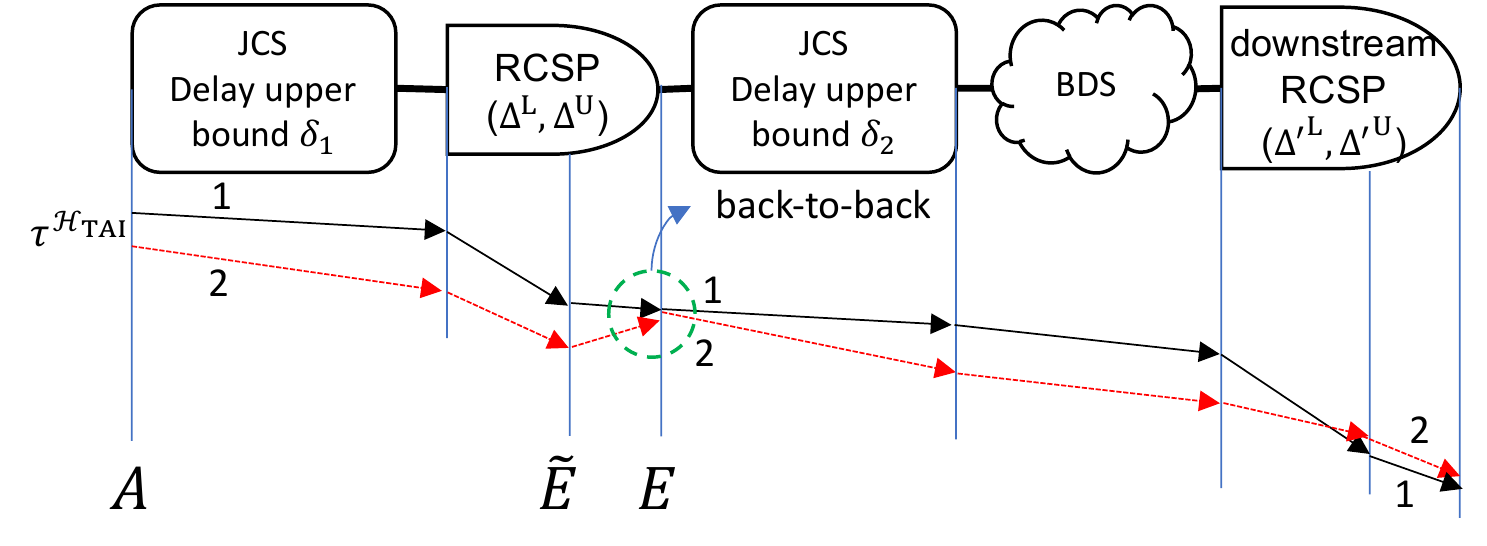}
	\caption{A scenario that two packets become back-to-back after an RCSP and reordering occurs in the downstream RCSP.}
	\label{fig:reordering-2}
\end{figure}

\textbf{Re-ordering example with RCSP.}
Consider Fig. \ref{fig:reordering-2} where the dampers are RCSP. Assume that the first JCS represents a router, the second JCS is a FIFO queue connected to a transmission line with a fixed rate and the BDS has zero jitter and represents a constant propagation delay; this resembles the first and second access routers in Example~2 of Section~\ref{sec:term}. Now, focus on the first JCS and the first RCSP. Suppose two packets~$1$ and~$2$ enter the JCS with interspacing $\tau^{\mathcal{H}_{\mathrm{TAI}}}$, measured in TAI (e.g. transmission time of packet $2$ from source, when packets are sent back-to-back from source), i.e.,
\begin{align}\label{eq:reordering-rcsp-1}
A^{\mathcal{H}_{\mathrm{TAI}}}_{2}-A^{\mathcal{H}_{\mathrm{TAI}}}_{1}=\tau^{\mathcal{H}_{\mathrm{TAI}}}.
\end{align}
Let $\kappa$ denote the delay difference of two packets from entrance of the  JCS to the theoretical eligibility time of the RCSP in TAI, i.e.,
\begin{align}
\kappa^{\mathcal{H}_{\mathrm{TAI}}}\triangleq\left(\tilde{E}^{\mathcal{H}_{\mathrm{TAI}}}_{1} - A^{\mathcal{H}_{\mathrm{TAI}}}_{1}\right)-\left(\tilde{E}^{\mathcal{H}_{\mathrm{TAI}}}_{2} - A^{\mathcal{H}_{\mathrm{TAI}}}_{2}\right).
\end{align}
$|\kappa^{\mathcal{H}_{\mathrm{TAI}}}|$ is less than the jitter bound of Theorem \ref{thm:perhop-concrete-delay} when $\Delta^{\mathrm{L}}=\Delta^{\mathrm{U}}=0$; $|\kappa^{\mathcal{H}_{\mathrm{TAI}}}|\leq V^0=\overline{\psi}+\underline{\psi}+2\epsilon$. Then by \eqref{eq:reordering-rcsp-1}, we have:
\begin{align}
\tilde{E}^{\mathcal{H}_{\mathrm{TAI}}}_{1}-\tilde{E}^{\mathcal{H}_{\mathrm{TAI}}}_{2} = \tau^{\mathcal{H}_{\mathrm{TAI}}}+ \kappa^{\mathcal{H}_{\mathrm{TAI}}},
\end{align}
which shows the interspacing between the theoretical eligibility times of two packets at the RCSP in TAI. Observing this interspacing with $\mathcal{H}_2$, we obtain
\begin{align}
\tilde{E}^{\mathcal{H}_{2}}_{1}-\tilde{E}^{\mathcal{H}_{2}}_{2} = \tau^{\mathcal{H}_{\mathrm{TAI}}}+ \kappa^{\mathcal{H}_{\mathrm{TAI}}}+\gamma,
\end{align}
where $\gamma$ is the difference in the measurement of the interspacing between $\mathcal{H}_{\mathrm{TAI}}$ and $\mathcal{H}_2$, bounded by \eqref{eq:gamma}. The actual eligibility times of the packets from RCSP are obtained by getting the floor of the theoretical eligibility times divided by $\Delta$ (Section \ref{sec:dm}). Hence, if $\tau^{\mathcal{H}_{\mathrm{TAI}}}+ \kappa^{\mathcal{H}_{\mathrm{TAI}}}+\gamma < \tolL$, the two packets may have the same actual eligibility times (leave RCSP back-to-back). The probability of this phenomenon is
\begin{align}
\nonumber \mathbb{P}\left[\mathrm{backToBack}\right] = 1-\frac{\tau^{\mathcal{H}_{\mathrm{TAI}}}+ \kappa^{\mathcal{H}_{\mathrm{TAI}}}+\gamma}{\tolL }
\approx 1-\frac{\tau^{\mathcal{H}_{\mathrm{TAI}}}}{\tolL },
\end{align}
which implies that, the smaller the interspacing of the two packets when entering the JCS, the larger is the probability of having the two packets with the same actual eligibility time and leave the RSCP back-to-back. When two packets are back-to-back from one RCSP, then similar to scenario 1, there is $50$\% chance of reordering for the two packets at the output of the next downstream RCSP. Due to the independence of the two events (being back-to-back at the output of the RCSP and reordering at next downstream RCSP), the chance of reordering is $0.5\left(1-{\tau^{\mathcal{H}_{\mathrm{TAI}}}}/\tolL\right)$.

One approach to tackle the reordering issue of dampers with tolerance is to place re-sequencing buffers after the dampers to correct the reordering that they cause. With this approach, it is crucial to find proper time-out value and size for the re-sequencing buffers. As shown in \cite{mohammadpour2020packet}, two reordering metrics, namely reordering late-time offset (RTO) and reordering byte offset (RBO) respectively give the time-out value and size of a re-sequencing buffer. We obtain these metrics for dampers as a direct result of \cite{mohammadpour2020packet}:

\begin{corollary}\label{col:perhop-rto-fixed}
	Consider Fig. \ref{fig:hop-concrete} and a flow that has arrival curve $\alpha^{\mathcal{H}_{\mathrm{TAI}}}$ at the entrance of the block. Then, the RTO for the flow from the entrance of the block to the output of the damper with tolerance, measured in TAI, is $\lambda^{\mathrm{TAI}}$ and the corresponding RBO is $\zeta$:
	\begin{align}
	\lambda^{\mathrm{TAI}} &= \left[V - \left(\alpha^{\mathcal{H}_{\mathrm{TAI}}}\right)^{\downarrow}(2L^{\min})\right]^+,\\
	\zeta &= \alpha^{\mathcal{H}_{\mathrm{TAI}}}\left(V\right) - L^{\min},
	\end{align}
	where $V$ is the jitter bound of the block, computed in Theorem \ref{thm:perhop-concrete-delay}, and $L^{\min}$ is the minimum packet length of the flow.
\end{corollary}

\begin{exmp}
	Consider Example 1 of Section \ref{sec:term} with the same assumptions made after Theorem \ref{thm:perhop-concrete-delay}. Suppose that a flow has leaky-bucket arrival curve with rate $16$~Mbps, in TAI, burstiness $10$~KBytes at source and minimum packet length $100$~Bytes. We computed that the jitter bound is $1.262~\mu$s from the source to the output of the first damper.  Then the RTO (time-out value) is $1.262~\mu$s and the RBO (required buffer size) is $10003$~Bytes.
\end{exmp}

Another approach to tackle reordering is to use dampers with FIFO constraint, as discussed in Section \ref{sec:dm} and analyzed in the next section.

\section{Analysis of Dampers with FIFO constraints} \label{sec:analysis-FIFO}
As mentioned earlier, one way to avoid packet reordering within dampers with tolerance is to replace them with dampers with FIFO constraint, namely, re-sequencing and HoL dampers. The goal of this section is to provide delay and jitter bounds when dampers with FIFO constraint are used. In this context, ``FIFO" and ``re-sequencing" are with respect to the aggregate of all packets that use a damper of interest. When all the BDSs and JCSs within a flow path are FIFO, using re-sequencing or HoL dampers, in contrary to dampers with tolerances, can provide end-to-end in-order packet delivery. However, this might impact the delay and jitter bounds computed in Theorem \ref{thm:perhop-concrete-delay}. To this end, we capture the impact of using re-sequencing or HoL dampers instead of dampers with tolerances in terms of delay and jitter bounds in Theorem \ref{thm:perhop-fifodamper-delay-fifo} and Theorem \ref{thm:perhop-jcats-delay-fifo} when all systems are FIFO. Then, we see in Theorem \ref{thm:perhop-fifodamper-delay-nonfifo} and Theorem \ref{thm:perhop-jcats-delay-nonfifo} that the presence of a non-FIFO system (BDS or JCS) in the flow path considerably worsens the delay and jitter bounds obtained when all systems are FIFO. This phenomenon does not occur with dampers without FIFO constraint because the results in Section~\ref{sec:analysis-noFIFO} hold whether the JCSs and BDSs are FIFO or not.

\begin{figure}[h]
	\centering
	\includegraphics[width= 0.8\linewidth]{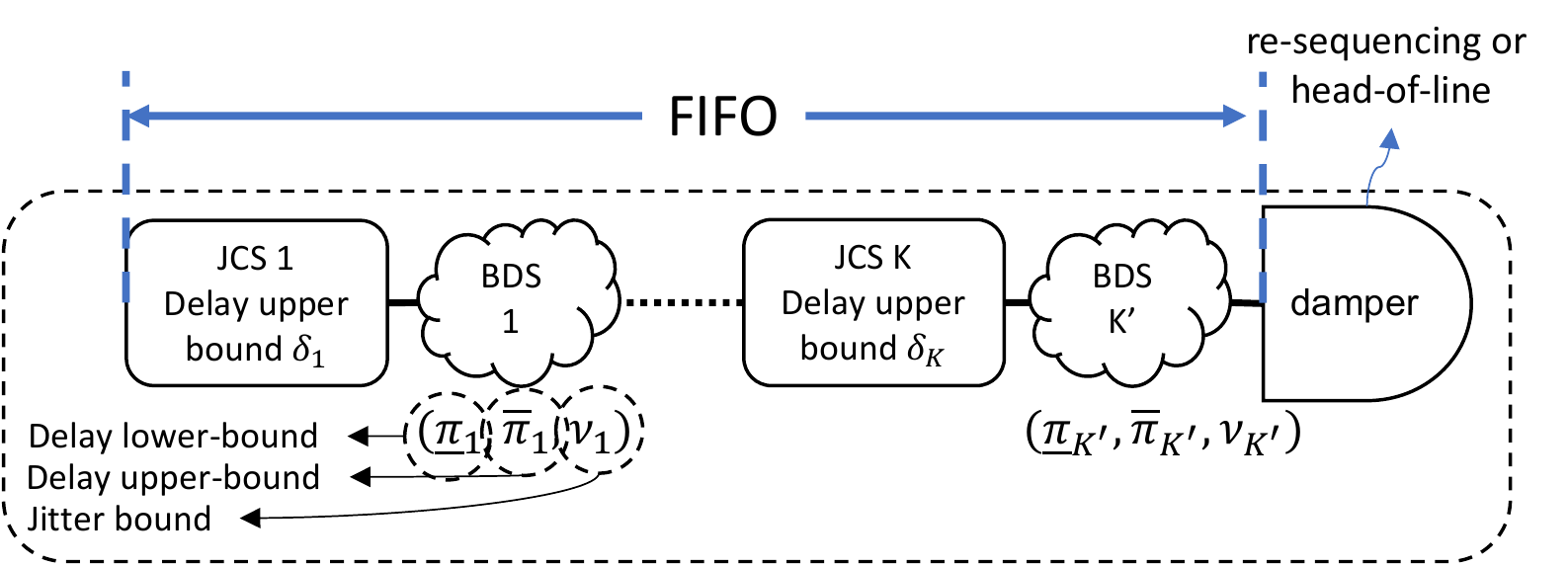}
	\caption{The notations used in Theorem \ref{thm:perhop-fifodamper-delay-fifo} and Theorem \ref{thm:perhop-jcats-delay-fifo}.}
	\label{fig:hop-fifo}
\end{figure}

\begin{theorem}\label{thm:perhop-fifodamper-delay-fifo}
	Consider the block of systems in Fig. \ref{fig:hop-fifo} where all the JCSs and BDSs are FIFO and the damper is an instance of re-sequencing dampers with tolerances $\left(\Delta^{\mathrm{L}},\Delta^{\mathrm{U}}\right)$. Assume that the clocks follow the description in Section \ref{sec:clock}.
	Then, the delay and jitter bounds of the block, in TAI, is the same as the bounds in Theorem \ref{thm:perhop-concrete-delay}.
\end{theorem}
The proof is in Appendix \ref{proof:perhop-fifodamper-delay-fifo}. It consists in two steps. First, we use an abstraction of a re-sequencing damper with tolerances $\left(\Delta^{\mathrm{L}},\Delta^{\mathrm{U}}\right)$ as a damper with tolerances $\left(\Delta^{\mathrm{L}},\Delta^{\mathrm{U}}\right)$ followed by a re-sequencing buffer that preserve the order of packet at their entrance to the damper with tolerances. Second, by the re-sequencing-for-free property of the re-sequencing buffers \cite{mohammadpour2020packet}, we obtain the bounds.
\begin{remark}
	We have seen in the previous section that even if all BDSs and JCSs are FIFO in a flow path, dampers with tolerance may cause packet reordering due to the tolerances, non-ideal clocks and errors in packet header computation. Theorem  \ref{thm:perhop-fifodamper-delay-fifo} indicates that in such a case, placing a re-sequencing damper avoids packet reordering with the same delay and jitter bounds as if dampers with tolerances are used.
\end{remark}

\begin{theorem}\label{thm:perhop-jcats-delay-fifo}
	Consider the block of systems in Fig. \ref{fig:hop-fifo} where all the JCSs and BDSs are FIFO and the damper is an instance of head-of-line dampers with tolerances ($\tolL,\tolU$) and processing-time bounds ($\phi^{\min},\phi^{\max}$). Assume that the clocks follow the description in Section \ref{sec:clock}.
	Then if $\phi^{\max}=0$, the delay and jitter bounds are the same as the bounds in Theorem \ref{thm:perhop-concrete-delay}. Otherwise, for a flow with per-packet arrival curve $\alpha$ at the entrance of the block,
	\begin{enumerate}
		\item the delay upper-bound is increased by  $\theta$,
		\item the delay lower-bound is increased by $\phi^{\min}$,
		\item the jitter bound is increased by $\theta-\phi^{\min}$,
	\end{enumerate}
where $\theta$ is a delay upper-bound of a single-server FIFO queue with maximum processing time of $\phi^{\max}$, computed as
\begin{align}\label{eq:theta}
\theta = \max_{k\in \mathbb{N}}\left\{k\phi^{\max}-\alpha^{\downarrow}(k)+ V \right\},
\end{align}
where $V$ is the jitter bound computed in Theorem \ref{thm:perhop-concrete-delay} and
$$\alpha^{\downarrow}(k) = \inf\left\{t\geq0| \alpha(t)\geq k\right\}.$$
\end{theorem}
The proof is in Appendix \ref{proof:perhop-jcats-delay-fifo}. The proof consists in two steps. First, we prove that an HoL damper is equivalent to re-sequencing damper with tolerances ($\tolL,\tolU$) followed by a single-server FIFO queue with service times within ($\phi^{\min},\phi^{\max}$). Second, using the bounds of Theorem \ref{thm:perhop-fifodamper-delay-fifo} and obtaining delay and jitter bounds on the single-server queue, the theorem is proven.

\begin{remark}
	HoL dampers, in contrary to re-sequencing dampers, imposes some queuing delay, captured by $\theta$ in \eqref{eq:theta}. The queuing delay is maximized for the last packet of a packet sequence when all become eligible at the same time; then since the HoL damper examines only the packet at the head of the queue, the last packet of the sequence is delayed as much as the processing delay of all the preceding packets.
\end{remark}

\begin{figure}[h]
	\centering
	\includegraphics[width= 0.9\linewidth]{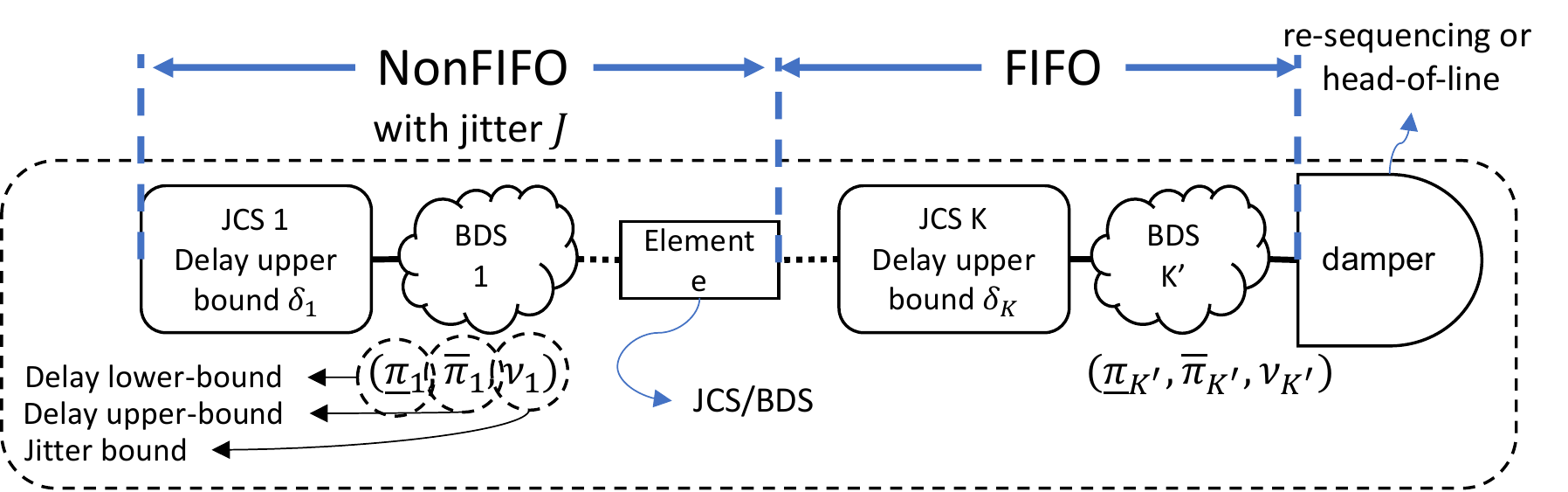}
	\caption{The notations used in Theorem \ref{thm:perhop-fifodamper-delay-nonfifo} and Theorem \ref{thm:perhop-jcats-delay-nonfifo}.}
	\label{fig:hop-jcats}
\end{figure}

So far we provide delay and jitter bounds when all the systems are FIFO; however, the FIFO condition might not be always met for all systems such as multi-stage switching fabrics, multi-path routing of packets or packet duplication \cite{bennett_packet_1999,laor_effect_2002,jaiswal_measurement_2007}.  Hence, in the following theorems we capture the impact of non-FIFO behavior of systems on the delay and jitter bounds when re-sequencing and HoL dampers are used.

\begin{theorem}\label{thm:perhop-fifodamper-delay-nonfifo}
	Consider Fig. \ref{fig:hop-jcats} where system $e$ (a BDS or a JCS) is the last non-FIFO system in the block and the damper is an instance of re-sequencing damper. Let us call $J$ as the jitter from JCS 1 to system $e$ (included), in TAI. Then, the delay upper-bound and jitter bound of the block, in TAI, are increased by $J$ comparing to the bounds in Theorem \ref{thm:perhop-fifodamper-delay-fifo}.
	
	 The bounds are tight, i.e., for any packet that experiences the delay equal to $\overline{D}$, there is system and an execution trace that another packet experiences a delay equal to $\overline{D}+J$.
\end{theorem}
The proof is in Appendix \ref{proof:perhop-fifodamper-delay-nonfifo}.  The proof has two parts. First, we show that delay upper-bound is increased by $J$ while the delay lower-bound remains unchanged. Second, we provide a scenario where two packets with interspacing $J$ enter the block and leave the element $e$ back-to-back while their order is changed. We show that when the second packet experiences a delay $\overline{D}$, the first packet experiences a delay of $\overline{D}+J$ and the second packet leaves the re-sequencing damper before the first packet.

\begin{theorem}\label{thm:perhop-jcats-delay-nonfifo}
	Consider Fig. \ref{fig:hop-jcats} where system $e$ (a BDS or a JCS) is the last non-FIFO system in the block and the damper is an instance of HoL damper. Let us call $J$ as the jitter from JCS~1 to system $e$ (included), in TAI. Then, comparing to Theorem \ref{thm:perhop-jcats-delay-fifo}, the delay upper-bound and jitter bound of the block, in TAI, are increased by $J$ if $\phi^{\max}=0$, and are increased by $2J$ if $\phi^{\max}>0$.
\end{theorem}
The proof is in Appendix \ref{proof:perhop-jcats-delay-nonfifo}. The proof consists in two steps. First, similarly to the proof of Theorem \ref{thm:perhop-fifodamper-delay-fifo}, we abstract an HoL damper as a re-sequencing damper followed by a single-server FIFO queue. Second, by summing the bounds obtained in Theorem \ref{thm:perhop-fifodamper-delay-nonfifo} and the bounds on the FIFO queue, the statement is proven. In the case $\phi^{\max}>0$, the bounds are increased once by $J$ within the re-sequencing damper and once within the FIFO queue as a result of propagated arrival curve at the output of the re-sequencing damper.

\begin{remark}
	Similarly to Corollary \ref{col:concrete-ac-prop}, propagated arrival curve of a flow, with arrival curve $\alpha^{\tai}(t)$ at the entrance of a block, is $\alpha^{\tai}(t+\bar{V})$ at the output of re-sequencing or HoL damper, where $\bar{V}$ is the jitter of the block computed by applying the corresponding theorem.
\end{remark}

\begin{remark}
	Theorem \ref{thm:perhop-fifodamper-delay-nonfifo} and Theorem \ref{thm:perhop-jcats-delay-nonfifo} show that when there is a non-FIFO system in a block, placement of a damper with FIFO constraint is counterproductive. First, comparing to placement of dampers with tolerances, the jitter is increased; in result, it leads to an increase in the burstiness of the propagated arrival curve. Second, the damper with FIFO constraint preserves the wrong order of the packets, which occurred within the non-FIFO system.
\end{remark}

\section{Numerical Evaluation}\label{sec:eval}
\begin{figure}[h]
	\centering
	\includegraphics[width= \linewidth]{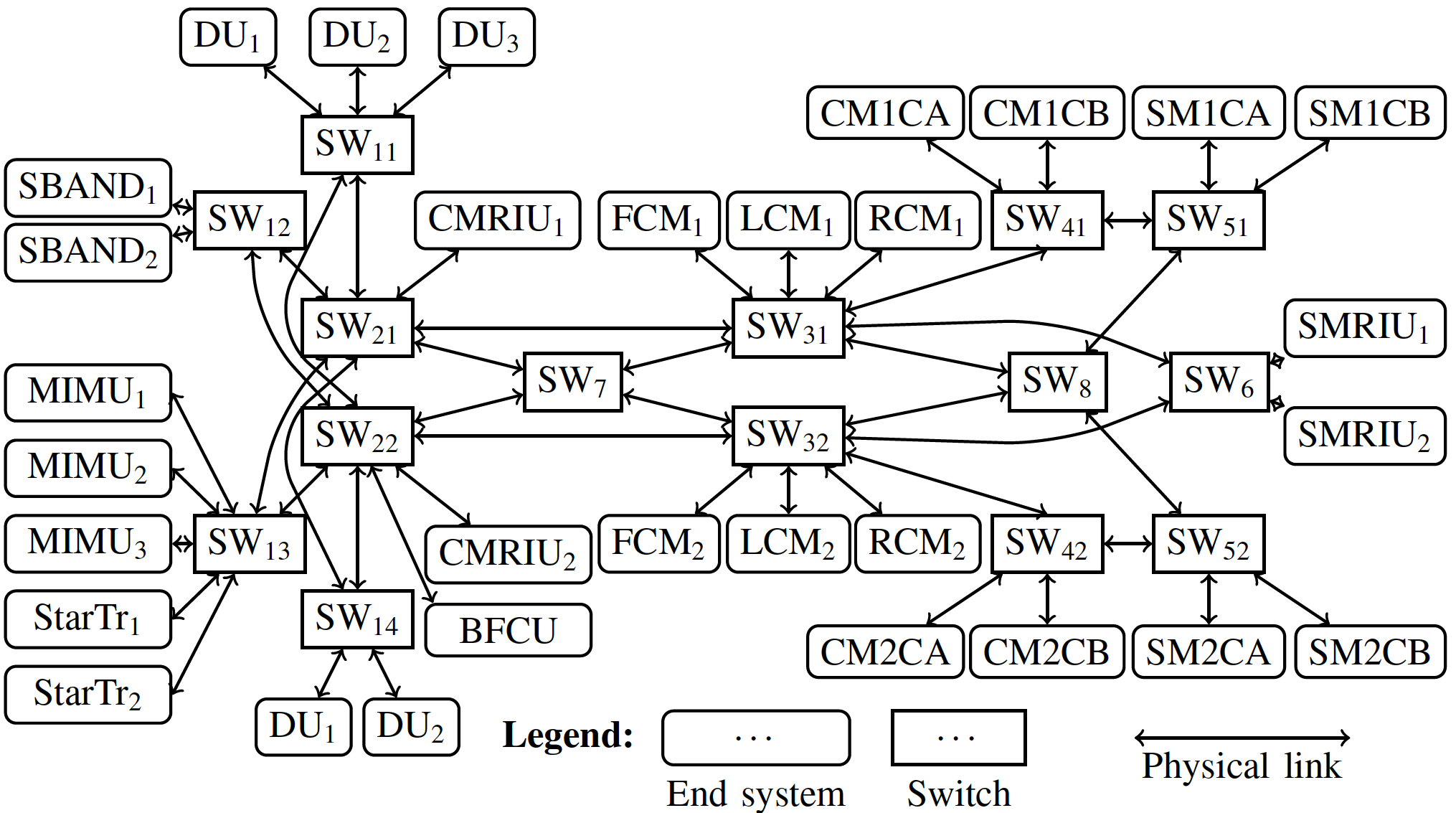}
	\caption{The Orion crew exploration vehicle network.}
	\label{fig:orion}
\end{figure}
We illustrate our theoretical results on the Orion crew exploration vehicle network, as described in \cite{obermaisser_time-triggered_2012} and depicted in Fig. \ref{fig:orion}, taken from \cite{thomas_on-cyclic_2019}. For the delay and jitter analysis, we used Fixed-Point TFA \cite{thomas_on-cyclic_2019,mifdaoui_beyond_2017} as there are cyclic dependencies. The device clocks are not synchronized.  The link rates are $1$~Gbps. The output ports use a non-preemptive TSN scheduler with Credit-based Shapers (CBSs) with per-class FIFO queuing \cite{mohammadpour_latency_2018,zhao_timing_2018}; from highest to lowest priority, the classes are CDT, A, B, and Best Effort (BE). The CBSs are used separately for classes A and B. The CBS parameters $\mathit{idleslope}$s are set to $50\%$ and $25\%$ of the link rate respectively for classes A and B \cite{zhao_timing_2018}. In each switch, the switching fabric has a delay between $0.5~\mu$s to $2~\mu$s \cite{nexus9508}. The CDT traffic has a leaky-bucket arrival curve with rate $6.4$ kilobytes per second and burst $64$~bytes. The maximum packet length of classes B and BE is $1500$~bytes. We focus on class A. Using the results in \cite{mohammadpour_latency_2018}, a rate-latency service curve offered to class A is $\beta(t) = 62.49e6[t-t_0]^+$~bytes with $t_0=12.5~\mu$s.

Class A contains $40$ flows with constant packet size $147$~bytes, which transmit $10$ packets every $8~$ms. The flows traverse between $2$ to $9$ hops. We assume all  switching fabrics and output queuing systems implement DHU unit and therefore are JCSs; the propagation delays are considered as BDSs with zero jitter. We examine the case where no damper is placed and the case where dampers are placed at every switch and the destinations. For the choice of dampers, we considered individually the full deployment of RCSP ($\Delta^{\mathrm{L}}=1~\mu$s, $\Delta^{\mathrm{U}}=2$~ns), RGCQ ($\Delta^{\mathrm{L}}=2$~ns, $\Delta^{\mathrm{U}}=1~\mu$s) with TE time-stamping, FOPLEQ ($\Delta^{\mathrm{L}}=1~\mu$s, $\Delta^{\mathrm{U}}=2$~ns) and a head-of-line damper ($\phi^{\min}=0, \phi^{\max}=5$~ns, $\tolL=\tolU=2$~ns).

\begin{figure}[h]
	\centering
	\includegraphics[width= \linewidth]{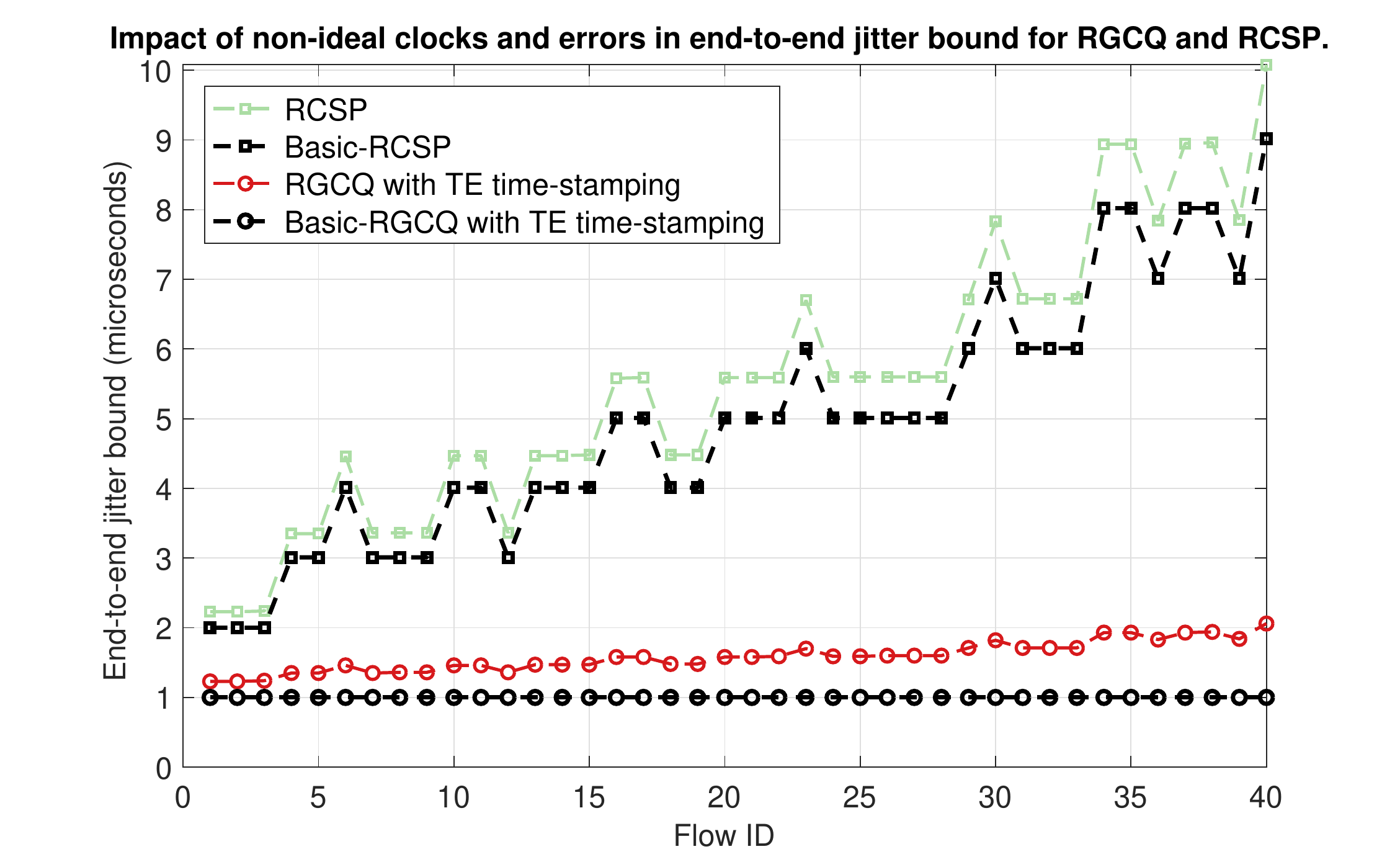}
	\caption{The end-to-end jitter bounds for RCSP and RGCQ with TE time-stamping using basic jitter computation and using theorems \ref{thm:perhop-concrete-delay} and \ref{thm:e2e-concrete}.}
	\label{fig:delay_jitter_orion_clocks}
\end{figure}
Fig. \ref{fig:delay_jitter_orion_clocks} shows the end-to-end jitter bounds of the flows for full deployment of RCSP and RGCQ with TE time-stamping. For each of the cases, the basic jitter computation only considers the jitter imposed by the tolerances of the dampers and ignore the impact of non-ideal clocks and errors in the computation of damper header. Fig. \ref{fig:delay_jitter_orion_clocks} also shows the true jitter bound for the case of RCSP, using Theorem \ref{thm:perhop-concrete-delay}, and for the case of RGCQ with TE time-stamping using Theorem \ref{thm:e2e-concrete}. We see that non-ideal clocks and errors can increase jitter by $11\%$ in the case of RCSP and $106\%$ in the case of RGCQ with TE time-stamping. We also see that the TE time-stamping used with RGCQ can significantly reduce the end-to-end jitter comparing to the default time-stamping used with RCSP.

\begin{figure*}[h]
	\centering
	\includegraphics[width= \textwidth]{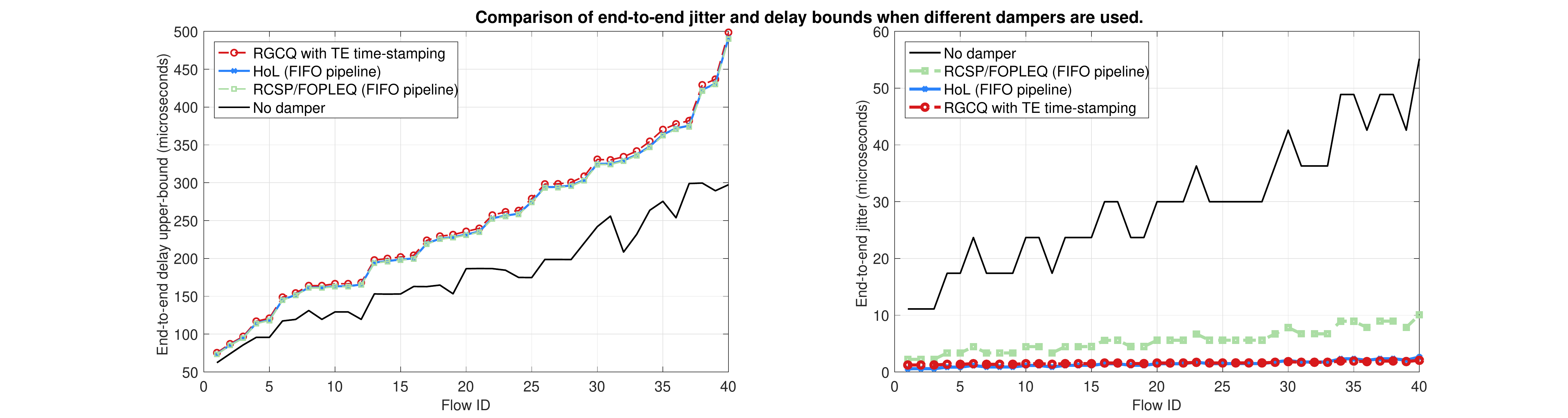}
	\caption{The end-to-end delay bounds (left) and jitter bounds (right) for full deployment of dampers and in the absence of dampers. For FOPLEQ and head-of-line (HoL) dampers, we assume that all the elements are FIFO.}
	\label{fig:delay_jitter_orion_fifo}
\end{figure*}
Fig. \ref{fig:delay_jitter_orion_fifo} shows the end-to-end delay and jitter bounds of the flows when no damper is used and when there is a full deployment as above. All switching fabrics are FIFO. We see that without damper, the delay upper-bound is smaller compared to full damper deployments; this is due to the line-shaping effect when computing the queuing delay bounds in the absence of dampers. However, as expected, the full deployment of dampers significantly reduces the jitter bounds. In this computation, the HoL damper provides quasi similar jitter bound as RGCQ with TE time-stamping and FOPLEQ gives the exact same jitter bound as RCSP as seen in Theorem~\ref{thm:perhop-fifodamper-delay-fifo}.

\begin{figure}[h]
	\centering
	\includegraphics[width= \linewidth]{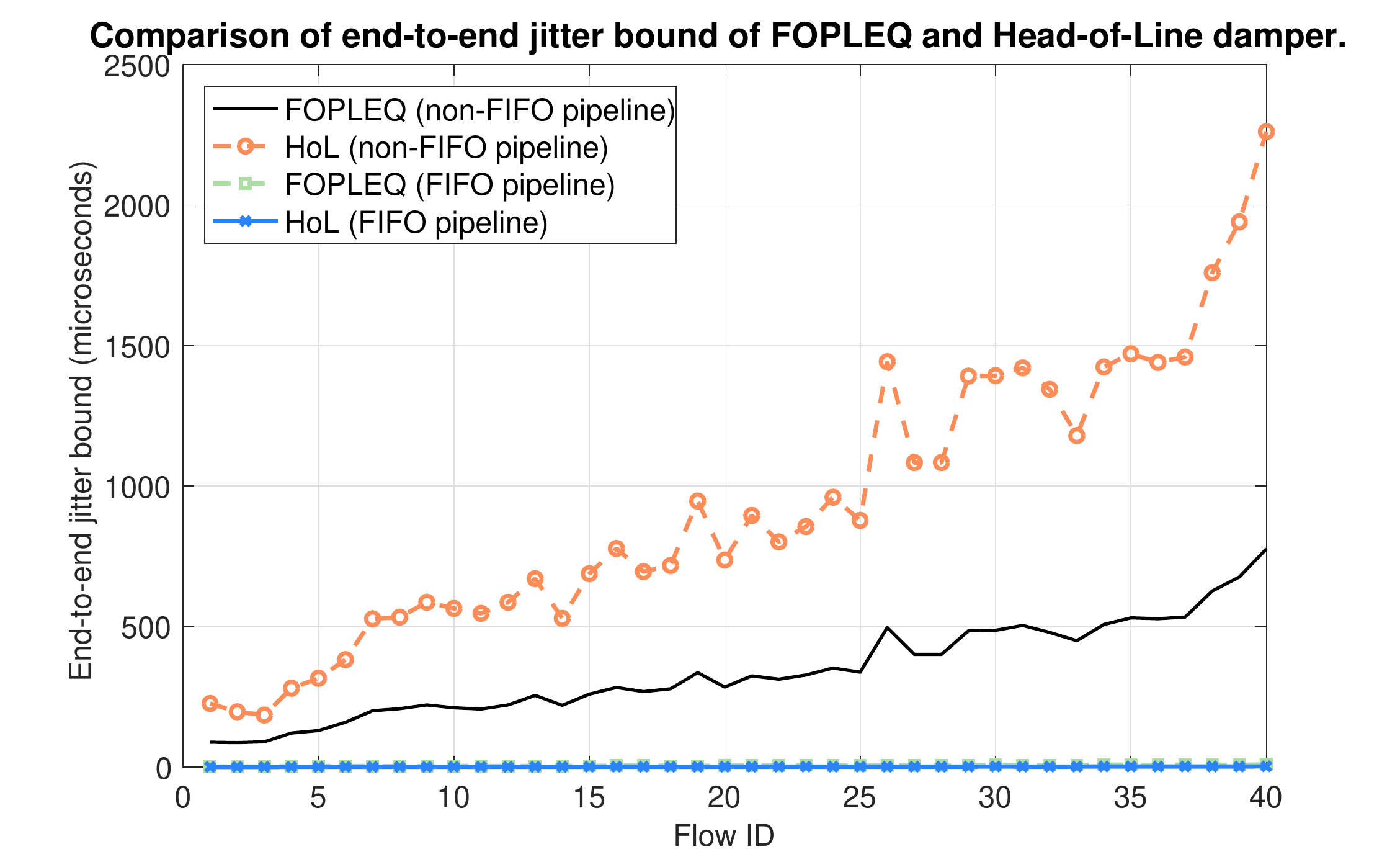}
	\caption{The end-to-end jitter bounds for FOPLEQ and head-of-line dampers when switching fabrics are FIFO or not.}
	\label{fig:delay_jitter_orion_nonfifo}
\end{figure}
Fig. \ref{fig:delay_jitter_orion_nonfifo} shows the end-to-end jitter bounds of the flows for FOPLEQ and HoL damper considering the switching fabrics are FIFO and are not FIFO. The figure shows that with FOPLEQ the jitter is significantly increased that is due to the jitter imposed by the output queuing, as seen in Theorem \ref{thm:perhop-fifodamper-delay-nonfifo}. It also shows that jitter bounds are worse in the case of HoL damper as discussed in Theorem \ref{thm:perhop-jcats-delay-nonfifo}. 
\section{Conclusion}\label{sec:conclusion}
We have presented a theory to compute delay and jitter bounds in a network that implements dampers with non-ideal clocks. We have shown that dampers without FIFO constraint can cause packet reordering even if all network elements are FIFO; re-sequencing dampers and head-of-line dampers avoid the problem; the former come with no jitter or delay penalty, and the latter with a small, quantified penalty.
However, when a flow path contains non-FIFO elements, re-sequencing dampers and head-of-line dampers do not perform well. 

%\section{Acknowledgments}
%This work was supported by Huawei Technologies Co., Ltd. in the framework of the project Large Scale Deterministic Network. The authors thank Bingyang Liu and Shoushou Ren for fruitful discussions. 

\bibliographystyle{IEEEtran}
\bibliography{ref}

\clearpage
\appendices
\twocolumn[
\begin{@twocolumnfalse}
	\begin{center}   
		{\Large\textbf{Supplementary Material}}\\
		{\textbf{Analysis of Dampers in Time-Sensitive Networks with Non-ideal Clocks}\\
			\textit{Ehsan Mohammadpour, Jean-Yves Le Boudec}}
	\end{center}
\end{@twocolumnfalse}
]

\section{Proofs}

\subsection{Lemmas for Section \ref{sec:dm}}\label{app:dm-extensive}

	\begin{lemma}\label{lem:rec-to-closed}
		Consider a non-decreasing sequence $I=\{I_1,\dots,I_n\}$ and a sequence $\phi=\{\phi_1,\dots,\phi_n\}$. Assume the sequence $O=\{O_1,\dots,O_n\}$ is defined by
		$$O_1=I_1+\phi_1,~~O_n = \max\left(I_n,O_{n-1}\right)+\phi_n.$$
		Then a closed-form formula for $O$ is:
		\begin{align}
		\nonumber O_n = \max_{m\leq n}\left\{I_m+\sum_{i=m}^{n}\phi_i\right\}.
		\end{align}
	\end{lemma}
\begin{proof}
	We prove by induction. Base case $n=1$.
	\begin{align}
	O_1=I_1+\phi_1,
	\end{align}
	as required by the lemma.
	
	Induction step. We assume that the lemma holds for all $i<n$. Then for $i=n-1$, by the closed-form formula we have:
	\begin{align}\label{eq:recur-proof-n-1}
	O_{n-1} = \max_{m\leq n-1}\left\{I_m+\sum_{i=m}^{n-1}\phi_i\right\}.
	\end{align}
	Then, using the recursive definition of $O_n$:
	\begin{align}
	\nonumber O_n &= \max\left(I_n,O_{n-1}\right)+\phi_n\\
	\nonumber &=\max\left(I_n+\phi_n,\max_{m\leq n-1}\left\{I_m+\sum_{i=m}^{n-1}\phi_i\right\}+\phi_n\right)\\
	\nonumber &=\max\left(I_n+\phi_n,\max_{m\leq n-1}\left\{I_m+\sum_{i=m}^{n}\phi_i\right\}\right)\\
	&=\max_{m\leq n}\left\{I_m+\sum_{i=m}^{n}\phi_i\right\}.
	\end{align}
\end{proof}

	\begin{lemma}\label{lem:hol-definition}
 Let $a$, $x^{\min}\leq x^{\max}$ and $y^{\min}\leq y^{\max}$ be five fixed real numbers. For any $z\in \Reals$, if
 \begin{equation}\label{eq:lem-hol-1}
   x^{\min}+ \max(a, y^{\min}) \leq z \leq x^{\max}+ \max(a, y^{\max}),
 \end{equation}
 then there exists some $x,y\in\Reals$ such that
 \begin{align}\label{eq:lem-hol-2a}
   z &=x + \max(a,y),  \\
   \label{eq:lem-hol-2b}x &\in [x^{\min}; x^{\max}], \\
   \label{eq:lem-hol-2c}y &\in [y^{\min}; y^{\max}].
 \end{align}
 Comversely, if there exists $x,y\in \Reals$ such that \eqref{eq:lem-hol-2a}--\eqref{eq:lem-hol-2c} hold, then $z$ satisfies \eqref{eq:lem-hol-1}.
	\end{lemma}
\begin{proof}
1) 	Let $D= [x^{\min}; x^{\max}]\times [y^{\min}; y^{\max}]$ and $f$ the mapping $D\to\Reals$ defined by $f(x,y)= x+ \max(a,y)$. $f$ is continuous and $D$ is compact and connected, therefore $f(D)$ is compact and connected. The compact and connected subsets of $\Reals$ are the closed, bounded intervals, therefore $f(D)=[m; M]$ for some $m,M\in\Reals$. Necessarily, $m$ is the minimum of $f$ over $D$ and $M$ is the maximum of $f$ over $D$.

Now for every $(x,y)\in D$:
\begin{equation}\label{eq:lem-hol-3}
  f(x^{\min},y^{\min})  \leq f(x,y)\leq f(x^{\max},y^{\max}).
\end{equation} It follows that the minimum of $f$ over $D$ is $f(x^{\min},y^{\min})$, i.e. $m=f(x^{\min},y^{\min})$. Similarly, $M=f(x^{\max},y^{\max})$.
Now let $z\in\Reals$ that satisfies \eqref{eq:lem-hol-1}, i.e. $z\in[m;M]=f(D)$. Thus, there exists some $(x,y)\in D$ such that $z=f(x,y)$, i.e.  there exists $x,y\in \Reals$ such that \eqref{eq:lem-hol-2a}--\eqref{eq:lem-hol-2c} hold.

2) Conversely, if there exists $x,y\in \Reals$ such that \eqref{eq:lem-hol-2a}--\eqref{eq:lem-hol-2c} hold, then $z=f(x,y)$ and thus $m\leq z\leq M$, i.e. \eqref{eq:lem-hol-1} holds.
\end{proof}

\subsection{Proof of Theorem~\ref{thm:perhop-concrete-delay}}\label{proof:perhop-concrete-delay}
 Let us define $d_i$ as the delay of JCS $i$ ($i=1,\dots,K$), $\pi_j$ as the delay of BDS $j$ ($j=1,\dots,K'$) and $t$ as the delay of the damper. Assume that JCS $i$ is operating with clock $\mathcal{H}_i$ and the damper is operating with clock $\mathcal{H}_{\mathrm{damper}}$. Then, the delay of the block, in TAI, is:
\begin{align}\label{eq:proof-concrete-damper-delay-2}
\nonumber d^{\mathcal{H}_{\mathrm{TAI}}}&= \sum_{i=1}^{K}d_{i}^{\mathcal{H}_{\mathrm{TAI}}}+\sum_{j=1}^{K'}\pi_j^{\mathcal{H}_{\mathrm{TAI}}} +t^{\mathcal{H}_{\mathrm{TAI}}}= \sum_{i=1}^{K}d_{i}^{\mathcal{H}_{i}}\\
&+\sum_{j=1}^{K'}\pi_j^{\mathcal{H}_{\mathrm{TAI}}} +t^{\mathcal{H}_{\mathrm{damper}}} + \gamma,
\end{align}
where $\gamma = \sum_{i=1}^{K}\left(d_{i}^{\mathcal{H}_{\mathrm{TAI}}}-d_{i}^{\mathcal{H}_i}\right)+\left(t^{\mathcal{H}_{\mathrm{TAI}}}-t^{\mathcal{H}_{\mathrm{damper}}}\right)$.
By \eqref{eq:elig-ideal-damper} and \eqref{eq:elig-concrete-damper}, the delay of a packet with damper header $H$ inside the damper is
\begin{align}\label{eq:proof-concrete-damper-delay-1}
H-\Delta^{\mathrm{L}} \leq t^{\mathcal{H}_{\mathrm{damper}}} \leq H+\Delta^{\mathrm{U}}.
\end{align}
Moreover, the damper header is incremented when the packet is passing by a JCS. Therefore, by \eqref{eq:earliness} and accounting for the errors in the computation of damper header, we have:
\begin{align}\label{eq:damper-header-concrete}
H &= \sum_{i=1}^{K}(\delta_i - d_i^{\mathcal{H}_{i}}+ e) = \sum_{i=1}^{K}\delta_i - \sum_{i=1}^{K} d_i^{\mathcal{H}_{i}}+ Ke,
\end{align}
where $e$ is defined in Section~\ref{sec:errors}. Using \eqref{eq:damper-header-concrete} and the upper bound in \eqref{eq:proof-concrete-damper-delay-1}, \eqref{eq:proof-concrete-damper-delay-2} gives:
\begin{align}
\nonumber d^{\mathcal{H}_{\mathrm{TAI}}}&\leq  \sum_{i=1}^{K}d_{i}^{\mathcal{H}_{i}} +\sum_{j=1}^{K'}\pi_j^{\mathcal{H}_{\mathrm{TAI}}} +\left(\sum_{i=1}^{K}\delta_i - \sum_{i=1}^{K} d_i^{\mathcal{H}_{i}}+ Ke\right) \\
&+\Delta^{\mathrm{U}}+ \gamma=\sum_{i=1}^{K}\delta_i+\sum_{j=1}^{K'}\pi_j^{\mathcal{H}_{\mathrm{TAI}}}+\Delta^{\mathrm{U}}+Ke+\gamma.
\end{align}
By Lemma \ref{lem:clock_perhop_concrete}, $\gamma \leq \overline{\psi}$; also $|e|\leq \epsilon$; furthermore, the delay of the packet inside the BDS $j$ is less than its worst-case delay, i.e., $\pi_{j,\mathrm{worst}}^{\mathcal{H}_{\mathrm{TAI}}}$.
%\jylb{$\mathrm{WCD}(\pi_j^{\mathcal{H}_{\mathrm{TAI}}})$ is improper notation. It means the result of applying the function $\mathrm{WCD}$ to the variable $\pi_j^{\mathcal{H}_{\mathrm{TAI}}}$.}
Therefore:
\begin{align}\label{eq:proof-concrete-damper-delay-3}
\nonumber d^{\mathcal{H}_{\mathrm{TAI}}}&\leq \sum_{i=1}^{K}\delta_i+\sum_{j=1}^{K'}\pi_{j,\mathrm{worst}}^{\mathcal{H}_{\mathrm{TAI}}}+\Delta^{\mathrm{U}}+K\epsilon+\overline{\psi}.
\end{align}

\noindent Similarly, using the lower bound in \eqref{eq:proof-concrete-damper-delay-1} and the value of $H$ in \eqref{eq:damper-header-concrete}, we have:
\begin{align}
\nonumber d^{\mathcal{H}_{\mathrm{TAI}}} &\geq \sum_{i=1}^{K}d_{i}^{\mathcal{H}_{i}} +\sum_{j=1}^{K'}\pi_j^{\mathcal{H}_{\mathrm{TAI}}} +\left(\sum_{i=1}^{K}\delta_i - \sum_{i=1}^{K} d_i^{\mathcal{H}_{i}}+ Ke\right) \\
&-\Delta^{\mathrm{L}}+ \gamma=\sum_{i=1}^{K}\delta_i+\sum_{j=1}^{K'}\pi_j^{\mathcal{H}_{\mathrm{TAI}}}-\Delta^{\mathrm{L}}+Ke+\gamma.
\end{align}
By Lemma \ref{lem:clock_perhop_concrete}, $\gamma \geq -\underline{\psi}$; also, $|e|\leq \epsilon$; furthermore, the delay of the packet inside the BDS $j$ is less than its best-case delay, i.e., $\pi_{j,\mathrm{best}}^{\mathcal{H}_{\mathrm{TAI}}}$. Therefore:
\begin{align}\label{eq:proof-concrete-damper-delay-4}
\nonumber d^{\mathcal{H}_{\mathrm{TAI}}} &\geq \sum_{i=1}^{K}\delta_i+\sum_{j=1}^{K'}\pi_{j,\mathrm{best}}^{\mathcal{H}_{\mathrm{TAI}}}-\Delta^{\mathrm{L}}-K\epsilon-\underline{\psi}_i.
\end{align}

By subtracting the lower and upper bounds in \eqref{eq:proof-concrete-damper-delay-3} and \eqref{eq:proof-concrete-damper-delay-4}, we obtain a jitter bound for the block:
\begin{align}
\nonumber V&=\sum_{j=1}^{K'}\left(\pi_{j,\mathrm{worst}}^{\mathcal{H}_{\mathrm{TAI}}}-\pi_{j,\mathrm{best}}^{\mathcal{H}_{\mathrm{TAI}}}\right)+\Delta^{\mathrm{U}}+\Delta^{\mathrm{L}}+2K\epsilon+\overline{\psi}\\
 &+\underline{\psi} = \sum_{j=1}^{K'} \nu_j^{\mathcal{H}_{\mathrm{TAI}}}+\Delta^{\mathrm{U}}+\Delta^{\mathrm{L}}+2K\epsilon+\overline{\psi}+\underline{\psi},
\end{align}
which proves the jitter bound. Note that $\pi_{j,\mathrm{worst}}^{\mathcal{H}_{\mathrm{TAI}}}-\pi_{j,\mathrm{best}}^{\mathcal{H}_{\mathrm{TAI}}}\leq \nu_j^{\mathcal{H}_{\mathrm{TAI}}}$.

\noindent Moreover, since $\pi_{j,\mathrm{worst}}^{\mathcal{H}_{\mathrm{TAI}}}\leq \overline{\pi}_j^{\mathcal{H}_{\mathrm{TAI}}}$ and $\pi_{j,\mathrm{best}}^{\mathcal{H}_{\mathrm{TAI}}}\geq \underline{\pi}_j^{\mathcal{H}_{\mathrm{TAI}}}$, we have:
\begin{align}
\nonumber d^{\mathcal{H}_{\mathrm{TAI}}} &\leq \sum_{i=1}^{K}\delta_i+\sum_{j=1}^{K'}\overline{\pi}_j^{\mathcal{H}_{\mathrm{TAI}}}+\Delta^{\mathrm{U}}+K\epsilon+\overline{\psi}=\overline{D},\\
d^{\mathcal{H}_{\mathrm{TAI}}} & \geq \sum_{i=1}^{K}\delta_i+\sum_{j=1}^{K'}\pi_j^{\mathcal{H}_{\mathrm{TAI}}}-\Delta^{\mathrm{L}}-K\epsilon-\underline{\psi}=\underline{D}.
\end{align}

\textbf{Proof of tightness.} Consider a packet that enters JCS $1$. We show scenarios where the packet reaches the delay bound in the statement of the theorem; since $\overline{\psi}$ can take different values due to the $\min(.)$ function, we give two different scenarios.

First scenario. Assume that the clocks $\mathcal{H}_i$, $i=1,...,K$, and $\mathcal{H}_{\mathrm{damper}}$ are adversarial and faster than TAI such that for any delay measurement $d$ we have:
\begin{align}\label{eq:proof-concrete-tight-1}
\nonumber d^{\mathcal{H}_{\mathrm{TAI}}} &= \rho d^{\mathcal{H}_i}+\eta,~~i=1,...,K\\
d^{\mathcal{H}_{\mathrm{TAI}}} &= \rho d^{\mathcal{H}_{\mathrm{damper}}}+\eta.
\end{align}
This scenario assumes that for any clock $\mathcal{H}_i$, $\rho d^{\mathcal{H}_i} +\eta \leq d^{\mathcal{H}_i}+2 \omega$ and $\rho d^{\mathcal{H}_{\mathrm{damper}}} +\eta \leq d^{\mathcal{H}_{\mathrm{damper}}}+2 \omega$.
%\jylb{This is possible only if we don't consider $\omega$.}
%Similarly, assume that clock $\mathcal{H}_{i+1}$ is adversarial and slower than TAI such that for any delay $d^{\mathcal{H}_{\mathrm{TAI}}}$, we have:
%$$d^{\mathcal{H}_{\mathrm{TAI}}} = \frac{1}{\rho}\left(d^{\mathcal{H}_{i+1}}-\eta\right).$$

Let us define $H_i$ as the damper header that is computed in JCS $i$. Then, let the packet experience a delay $d_i^{\mathcal{H}_i}\leq \delta_i$ in JCS $i$ in its local time; then the damper header written for this packet is
$$H_1=\delta_1-d_1^{\mathcal{H}_1}+\epsilon,$$
$$H_i=H_{i-1}+\delta_i-d_i^{\mathcal{H}_i}+\epsilon, ~~~i=2,...,K$$
 assuming adversarial condition in the damper header computation error. The packet experiences a delay equal to $\overline{\pi}_j^{\mathcal{H}_{\mathrm{TAI}}}$ in BDS $j$. Finally, the damper computes the eligibility time and releases it at the latest; i.e., the packet experiences a delay
 $$t^{\mathcal{H}_{\mathrm{damper}}}=H_K+\Delta^{\mathrm{U}}=\sum_{u=1}^{K}\delta_u-\sum_{u=1}^{K}d_u^{\mathcal{H}_u}+K\epsilon+\Delta^{\mathrm{U}}.$$
 Therefore, for the damper, the delay of the packet in TAI, using \eqref{eq:proof-concrete-tight-1}, is:
 \begin{align}
 t^{\mathcal{H}_{\mathrm{TAI}}} = \rho\left(\sum_{u=1}^{K}\delta_u-\sum_{u=1}^{K}d_u^{\mathcal{H}_u}+K\epsilon+\Delta^{\mathrm{U}}\right)+\eta.
 \end{align}
 Also, for JCS $i$, the delay of the packet in TAI is:
 \begin{align}
 d_i^{\mathcal{H}_{\mathrm{TAI}}} = \rho d_i^{\mathcal{H}_i} + \eta.
 \end{align}
Now, to compute the per-hop delay of the packet, we sum up all the delays in TAI:
\begin{align}
\nonumber d_{\mathrm{hop}}^{\mathcal{H}_{\mathrm{TAI}}} &= \sum_{u=1}^{K}d_u^{\mathcal{H}_{\mathrm{TAI}}}+\sum_{j=1}^{K'}\overline{\pi}_j^{\mathcal{H}_{\mathrm{TAI}}} +t^{\mathcal{H}_{\mathrm{TAI}}} = \sum_{u=1}^{K}\left(\rho d_u^{\mathcal{H}_i} + \eta \right)\\
\nonumber &+\sum_{j=1}^{K'}\overline{\pi}_j^{\mathcal{H}_{\mathrm{TAI}}}+\rho\left(\sum_{u=1}^{K}\delta_u-\sum_{u=1}^{K}d_u^{\mathcal{H}_u}+K\epsilon+\Delta^{\mathrm{U}}\right)+\eta\\
\nonumber&= \rho (\sum_{u=1}^{K}\delta_u+K\epsilon+\Delta^{\mathrm{U}}) +\sum_{j=1}^{K'}\overline{\pi}_j^{\mathcal{H}_{\mathrm{TAI}}}+ (K+1)\eta\\
&=\sum_{u=1}^{K}\delta_u+\sum_{j=1}^{K'}\overline{\pi}_j^{\mathcal{H}_{\mathrm{TAI}}}+\Delta^{\mathrm{U}}+K\epsilon+\overline{\psi},
\end{align}
which is equal to the delay bound in the statement of the theorem.

Second scenario. Assume that the clocks $\mathcal{H}_i$, $i=1,...,K$, and $\mathcal{H}_{\mathrm{damper}}$ are adversarial and faster than TAI such that for any delay measurement $d$ we have:
\begin{align}\label{eq:proof-concrete-tight-2}
\nonumber d^{\mathcal{H}_{\mathrm{TAI}}} &= d^{\mathcal{H}_i}+2 \omega,~~i=1,...,K\\
d^{\mathcal{H}_{\mathrm{TAI}}} &= d^{\mathcal{H}_{\mathrm{damper}}}+2 \omega.
\end{align}
This scenario assumes that for any clock $\mathcal{H}_i$, $\rho d^{\mathcal{H}_i} +\eta \geq d^{\mathcal{H}_i}+2 \omega$ and $\rho d^{\mathcal{H}_{\mathrm{damper}}} +\eta \geq d^{\mathcal{H}_{\mathrm{damper}}}+2 \omega$. Then following the same steps as the previous scenario, we have:
\begin{align}
\nonumber t^{\mathcal{H}_{\mathrm{TAI}}} &= \sum_{u=1}^{K}\delta_u-\sum_{u=1}^{K}d_u^{\mathcal{H}_u}+K\epsilon+\Delta^{\mathrm{U}} + 2\omega,\\
d_i^{\mathcal{H}_{\mathrm{TAI}}} &= d_i^{\mathcal{H}_i} + 2 \omega,~~~\forall i\in\{1,\dots,K\}.
\end{align}
This gives:
\begin{align}
\nonumber d_{\mathrm{hop}}^{\mathcal{H}_{\mathrm{TAI}}} &= \sum_{u=1}^{K}d_u^{\mathcal{H}_{\mathrm{TAI}}}+\sum_{j=1}^{K'}\overline{\pi}_j^{\mathcal{H}_{\mathrm{TAI}}} +t^{\mathcal{H}_{\mathrm{TAI}}} = \sum_{u=1}^{K}\left( d_u^{\mathcal{H}_i} +2\omega\right)\\
\nonumber &+\sum_{j=1}^{K'}\overline{\pi}_j^{\mathcal{H}_{\mathrm{TAI}}}+\sum_{u=1}^{K}\delta_u-\sum_{u=1}^{K}d_u^{\mathcal{H}_u}+K\epsilon+\Delta^{\mathrm{U}} + 2\omega\\
\nonumber&= \sum_{u=1}^{K}\delta_u+K\epsilon+\Delta^{\mathrm{U}} +\sum_{j=1}^{K'}\overline{\pi}_j^{\mathcal{H}_{\mathrm{TAI}}}+ 2(K+1)\omega\\
&=\sum_{u=1}^{K}\delta_u+\sum_{j=1}^{K'}\overline{\pi}_j^{\mathcal{H}_{\mathrm{TAI}}}+\Delta^{\mathrm{U}}+K\epsilon+\overline{\psi},
\end{align}
which is equal to the delay bound in the statement of the theorem. The tightness for delay lower bound happens when the clocks $\mathcal{H}_i$, $i=1,...,K$, and $\mathcal{H}_{\mathrm{damper}}$ are adversarial and slower than TAI. Similarly to the tightness proof of delay upper-bound, two tightness scenarios are given for the two possible values of $\underline{\psi}$. For the first scenario, for any delay measurement $d$, we have:
\begin{align}
\nonumber d^{\mathcal{H}_{\mathrm{TAI}}} &= \frac{1}{\rho}\left(d^{\mathcal{H}_i}-\eta\right),~~i=1,...,K,\\
d^{\mathcal{H}_{\mathrm{TAI}}} &= \frac{1}{\rho}\left(d^{\mathcal{H}_{\mathrm{damper}}}-\eta\right),
\end{align}
and for the second scenario, we have:
\begin{align}
\nonumber d^{\mathcal{H}_{\mathrm{TAI}}} &= d^{\mathcal{H}_i}-2\omega,~~i=1,...,K,\\
d^{\mathcal{H}_{\mathrm{TAI}}} &= d^{\mathcal{H}_{\mathrm{damper}}}-2\omega.
\end{align}
Considering the damper releases the packets at the earliest, the rest of the proof follows the same steps as the tightness proof of delay upper-bound.

\begin{lemma} \label{lem:clock_perhop_concrete}
	$-\underline{\psi} \leq \gamma\leq \overline{\psi}$, where $\underline{\psi}$ and $\overline{\psi}$ are defined in \eqref{eq:psi}:
\end{lemma}
\begin{proof}
	By definition of $\gamma$, we have:
	\begin{align}
	\gamma &= \sum_{j=1}^{K}\left(d_{j}^{\mathcal{H}_{\mathrm{TAI}}}-d_{j}^{\mathcal{H}_j}\right)+\left(t^{\mathcal{H}_{\mathrm{TAI}}}-t^{\mathcal{H}_{\mathrm{damper}}}\right).
	\end{align}
	First we prove the upper bound. By \eqref{eq:gamma} the following holds for any JCS $j$:
	\begin{align}\label{eq:proof-clock-up-1}
	\nonumber&d_{j}^{\mathcal{H}_{\mathrm{TAI}}}-d_{j}^{\mathcal{H}_j} \leq (\rho-1)d_{j}^{\mathcal{H}_{j}}+\eta,\\
	& d_{j}^{\mathcal{H}_{\mathrm{TAI}}}-d_{j}^{\mathcal{H}_j} \leq 2\omega.
	\end{align}
	 Using the value of $H$ in \eqref{eq:damper-header-concrete} and the upper bound in \eqref{eq:proof-concrete-damper-delay-1}, we have $t^{\mathcal{H}_{\mathrm{damper}}} \leq \sum_{j=1}^{K}\left(\delta_j-d_{j}^{\mathcal{H}_{j}}+\epsilon\right)+\Delta^{\mathrm{U}}$. Hence, similarly to the previous equation:
	\begin{align}\label{eq:proof-clock-up-2}
	\nonumber& t^{\mathcal{H}_{\mathrm{TAI}}}-t^{\mathcal{H}_{\mathrm{damper}}} \leq (\rho-1)\Big(\Delta^{\mathrm{U}}+\sum_{j=1}^{K}\left(\delta_j-d_{j}^{\mathcal{H}_{j}}+\epsilon\right)\Big)+\eta,\\
	& t^{\mathcal{H}_{\mathrm{TAI}}}-t^{\mathcal{H}_{\mathrm{damper}}} \leq 2\omega.
	\end{align}
	We first consider the case that the synchronization inequality is dominating for all systems (i.e., the second line of \eqref{eq:proof-clock-up-1} and \eqref{eq:proof-clock-up-2}). Hence, we have:
	\begin{align}\label{eq:gamma-up-1}
	\gamma \leq \sum_{j=1}^{K} 2\omega + 2\omega = 2(K+1) \omega.
	\end{align}
	Second, we consider the case that the free-running mode is dominating for all systems (i.e., in the first line of \eqref{eq:proof-clock-up-1} and \eqref{eq:proof-clock-up-2}). Then, we have:
	\begin{align}\label{eq:gamma-up-2}
	\nonumber\gamma \leq &\sum_{j=1}^{K} \left((\rho-1)d_{j}^{\mathcal{H}_{j}}+\eta\right) + (\rho-1)\Delta^{\mathrm{U}}\\
	\nonumber &+(\rho-1)\left(\sum_{j=1}^{K}\left(\delta_j-d_{j}^{\mathcal{H}_{j}}+\epsilon\right)\right)+\eta\\
	&= (\rho-1)\left(\Delta^{\mathrm{U}}+\sum_{j=1}^{K}(\delta_j+\epsilon)\right)+ (K+1)\eta.
	\end{align}
	Finally, by \eqref{eq:gamma-up-1} and \eqref{eq:gamma-up-2}, $\gamma\leq \overline{\psi}$.
	
	\noindent Next, we prove the lower bound. Similarly, by \eqref{eq:gamma} the following holds for any JCS $j$:
	\begin{align}\label{eq:proof-clock-low-1}
	\nonumber&d_{j}^{\mathcal{H}_{\mathrm{TAI}}}-d_{j}^{\mathcal{H}_j} \geq -\left(1-\frac{1}{\rho}\right)d_{j}^{\mathcal{H}_{j}}-\frac{\eta}{\rho},\\
	& d_{j}^{\mathcal{H}_{\mathrm{TAI}}}-d_{j}^{\mathcal{H}_j} \geq -2\omega.
	\end{align}
	Using the value of $H$ in \eqref{eq:damper-header-concrete} and the upper bound in \eqref{eq:proof-concrete-damper-delay-1}, we have $t^{\mathcal{H}_{\mathrm{damper}}} \geq \sum_{j=1}^{K}\left(\delta_j-d_{j}^{\mathcal{H}_{j}}-\epsilon\right)-\Delta^{\mathrm{L}}$. Hence:
	\begin{align}\label{eq:proof-clock-low-2}
	\nonumber& t^{\mathcal{H}_{\mathrm{TAI}}}-t^{\mathcal{H}_{\mathrm{damper}}} \geq -(1-\frac{1}{\rho})(-\Delta^{\mathrm{L}}+\sum_{j=1}^{K}(\delta_j-d_{j}^{\mathcal{H}_{j}}-\epsilon))\\
	\nonumber &-\frac{\eta}{\rho},\\
	& t^{\mathcal{H}_{\mathrm{TAI}}}-t^{\mathcal{H}_{\mathrm{damper}}} \geq -2\omega.
	\end{align}
	We first consider the case that the synchronization inequality is dominating for all systems (i.e., the second line of \eqref{eq:proof-clock-up-1} and \eqref{eq:proof-clock-up-2}). Hence, we have:
	\begin{align}\label{eq:gamma-low-1}
	\gamma \geq -\sum_{j=1}^{K} 2\omega - 2\omega = -2(K+1) \omega.
	\end{align}
	Second, we consider the case that the free-running mode is dominating for all systems (i.e., in the first line of \eqref{eq:proof-clock-up-1} and \eqref{eq:proof-clock-up-2}). Then, we have:
	\begin{align}\label{eq:gamma-low-2}
	\nonumber\gamma \geq & -\left(1-\frac{1}{\rho}\right)\sum_{j=1}^{K}d_{j}^{\mathcal{H}_{j}}-\frac{K\eta}{\rho}\\
	\nonumber&-\left(1-\frac{1}{\rho}\right)\left(-\Delta^{\mathrm{L}}+\sum_{j=1}^{K}\left(\delta_j-d_{j}^{\mathcal{H}_{j}}-\epsilon\right)\right)-\frac{\eta}{\rho} \\
	&= -\left(1-\frac{1}{\rho}\right)\left(-\Delta^{\mathrm{L}}+\sum_{j=1}^{K}\left(\delta_j-\epsilon\right)\right)-\frac{(K+1)\eta}{\rho}.
	\end{align}
	Finally, by \eqref{eq:gamma-low-1} and \eqref{eq:gamma-low-2}, $\gamma\geq -\underline{\psi}$.
\end{proof}

\subsection{Proof of Theorem~\ref{thm:e2e-concrete}}\label{proof:e2e-concrete}
%%%%%%%%%%%%%%%%%%%%%%%%%%%%%%%%%%
%\begin{figure}[h]
%	\centering
%	\includegraphics[width= \linewidth]{fig/concrete-e2e-proof.pdf}
%	\caption{The notations used in the proof of Theorem \ref{thm:e2e-concrete}.}
%	\label{fig:e2e-concrete-proof}
%\end{figure}
Consider a packet that enters block $1$ at time $A$. Let us denote the theoretical and actual eligibility time of the packet from the damper of block $i$ as $\tilde{E}_i$ and $E_i$ respectively. Then, the delay from $A$ to the output of block $N$, i.e., $E_N$, is:
\begin{align}\label{eq:proof-e2e-te-1}
\nonumber d_{\mathrm{e2e}}^{\mathcal{H}_{\mathrm{TAI}}} &= E_N^{\mathcal{H}_{\mathrm{TAI}}}-A^{\mathcal{H}_{\mathrm{TAI}}} = \left(\tilde{E}_1^{\mathcal{H}_{\mathrm{TAI}}}-A^{\mathcal{H}_{\mathrm{TAI}}}\right) +\\
& \sum_{i=2}^{N-1}\left(\tilde{E}_{i}^{\mathcal{H}_{\mathrm{TAI}}}-\tilde{E}_{i-1}^{\mathcal{H}_{\mathrm{TAI}}}\right)+\left({E}_{N}^{\mathcal{H}_{\mathrm{TAI}}}-\tilde{E}_{N-1}^{\mathcal{H}_{\mathrm{TAI}}}\right).
\end{align}
By Theorem \ref{thm:perhop-concrete-delay} and setting the tolerances to zero, we can find delay and jitter bounds for $\left(\tilde{E}_1^{\mathcal{H}_{\mathrm{TAI}}}-A^{\mathcal{H}_{\mathrm{TAI}}}\right)$ in TAI. For any term $\left(\tilde{E}_{i}^{\mathcal{H}_{\mathrm{TAI}}}-\tilde{E}_{i-1}^{\mathcal{H}_{\mathrm{TAI}}}\right)$ in the summation, using Lemma \ref{lem:perhop-te}, we can obtain delay and jitter bound by setting $\Delta_i^{\mathrm{L}}=\Delta_i^{\mathrm{U}}=0$ (since we are interested to find the delay to the theoretical eligibility time at the damper $i$). Finally, we can apply again Lemma \ref{lem:perhop-te} to get the bounds for $\left({E}_{N}^{\mathcal{H}_{\mathrm{TAI}}}-\tilde{E}_{N-1}^{\mathcal{H}_{\mathrm{TAI}}}\right)$. By summing up the delay and jitter bounds of each term in \eqref{eq:proof-e2e-te-1}, we get the bounds in the statement of the theorem.
\begin{lemma}\label{lem:perhop-te}
	Consider a block $i$ in Fig. \ref{fig:e2e-concrete} that has $K_i$ JCSs. Assume that TE time-stamping is used to compute damper headers. Let $\delta_{\mathrm{blk}_i}$ denote the sum of the delay bounds of the JCSs in the block, $\underline{\pi}_{\mathrm{blk}_i}^{\mathcal{H}_{\mathrm{TAI}}},\overline{\pi}_{\mathrm{blk}_i}^{\mathcal{H}_{\mathrm{TAI}}}$ and $\nu_{\mathrm{blk}_i}^{\mathcal{H}_{\mathrm{TAI}}}$ respectively denote the sum of delay lower and upper bounds and jitter bound of the BDSs. Then, the delay of a packet from theoretical eligibility time of damper $i-1$, $i=2,\dots,N$, to the actual eligibility time of damper $i$, in TAI, is upper-bounded by $\overline{D}_i$, low-bounded by $\underline{D}_i$ and has jitter bound $V_i$:\\
	\begin{align}
	\nonumber\overline{D}_i&=\delta_{\mathrm{blk}_i}+\overline{\pi}_{\mathrm{blk}_i}^{\mathcal{H}_{\mathrm{TAI}}}+\Delta_{i-1}^{\mathrm{U}}+\Delta_{i}^{\mathrm{U}}+K_i\epsilon +\overline{\psi}'_i,\\
	\nonumber \underline{D}_i&=\delta_{\mathrm{blk}_i}+\underline{\pi}_{\mathrm{blk}_i}^{\mathcal{H}_{\mathrm{TAI}}}+\Delta_{i-1}^{\mathrm{U}} -\Delta_i^{\mathrm{L}}-K_i\epsilon -\underline{\psi}'_i,\\
	\nonumber V_i&=\nu_{\mathrm{blk}_i}^{\mathcal{H}_{\mathrm{TAI}}}  + \Delta_i^{\mathrm{U}}+\Delta_i^{\mathrm{L}} + 2K_i\epsilon+\overline{\psi}_i+\underline{\psi}'_i,
	\end{align}
	where,
	\begin{align}\label{eq:psi'}
	\nonumber \overline{\psi}_i'&=\min\Big((\rho-1)\left(\delta_{\mathrm{blk}_i}+\Delta_{i-1}^{\mathrm{U}}+\Delta_{i}^{\mathrm{U}}+K_i\epsilon\right)+ (K_i+1)\eta\\
	\nonumber&,2(K_i+1) \omega\Big),\\
	\nonumber\underline{\psi}'_i &= \min\Bigg(\left(1-\frac{1}{\rho}\right)\left(\delta_{\mathrm{blk}_i}+\Delta_{i-1}^{\mathrm{U}}-\Delta_i^{\mathrm{L}}-K_i\epsilon\right)\\
	&+\frac{(K_i+1)\eta}{\rho},2(K_i+1) \omega\Bigg).
	\end{align}
\end{lemma}
\begin{proof}
	The proof follows the same as Theorem \ref{thm:perhop-concrete-delay} and using:
	\begin{align}
	\nonumber H &= (\delta_1 +\Delta_{i-1}^{\mathrm{U}}  - d_1^{\mathcal{H}_{1}}+ e)+\sum_{j=2}^{K}(\delta_j -d_j^{\mathcal{H}_{j}} + e)\\
	&= \sum_{j=1}^{K}\delta_j - \sum_{i=1}^{K} d_i^{\mathcal{H}_{i}}+\Delta_{i-1}^{\mathrm{U}}+ Ke,
	\end{align}
	instead of \eqref{eq:damper-header-concrete} for damper header computation. Note that here $d_j^{\mathcal{H}_{j}}$ is the delay from of the packet from start of time stamping at JCS $j$ to its departure time.
\end{proof}

\subsection{Proof of Theorem~\ref{thm:perhop-fifodamper-delay-fifo}}\label{proof:perhop-fifodamper-delay-fifo}
\begin{figure}[h]
	\centering
	\includegraphics[width=0.7 \linewidth]{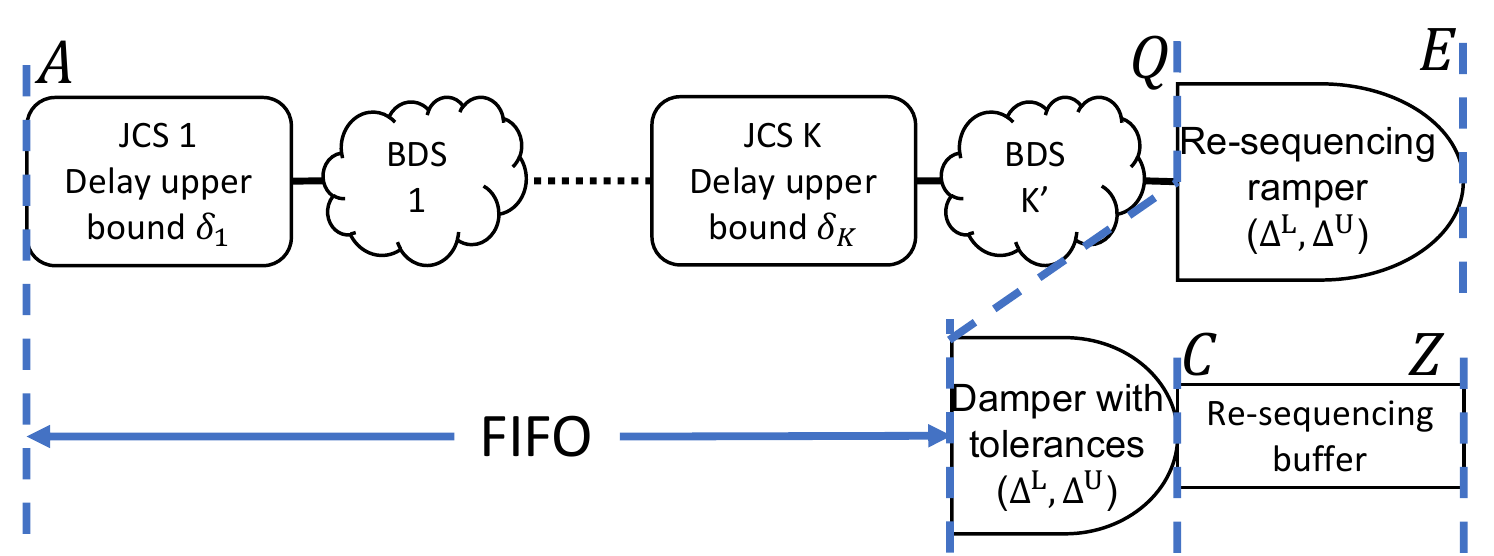}
	\caption{The notations used in the proof of Theorem \ref{thm:perhop-fifodamper-delay-fifo}.}
	\label{fig:fifodamper-fifo-reseq}
\end{figure}
Consider Fig. \ref{fig:fifodamper-fifo-reseq} where $A$ is the sequence of packet arrival times at  entrance of the block, $Q$ is the sequence of arrivals to the re-sequencing damper, and $E$ is the sequence of actual eligibility times at the re-sequencing damper. By definition, the re-sequencing damper behaves as a damper with tolerances ($\Delta^{\mathrm{L}},\Delta^{\mathrm{U}}$) followed by a re-sequencing buffer.
%By Lemma \ref{lem:fifodamper-concrete-reseq}, the re-sequencing damper is equivalent to a lossless concrete damper followed by a re-sequencing buffer.
Denote with $C$ the actual eligibility times from the damper with tolerance and with $Z$ the output times of the re-sequencing buffer; the equivalence means that $E=Z$.

By Theorem \ref{thm:perhop-concrete-delay}, the delay from $A$ to $C$ has lower bound $\underline{D}$, upper bound $\overline{D}$ and the jitter bound is $V$.
Now, let $\mathcal{P}$ and $\mathcal{P'}$ be the sets of packets seen respectively at $A$ and $Q$, i.e., $\mathcal{P'}\subseteq\mathcal{P}$; by defining $\mathcal{P'}$, we only include the packets of $\mathcal{P}$ that arrive to $Q$; therefore, by \cite[Theorem 4]{mohammadpour2020packet}, the re-sequencing buffer does not increase the worst-case delay and jitter from $A$ to $C$. Hence, the delay and jitter bounds from $A$ to $Z$ are the same as from $A$ to $C$, which proves the theorem.

\subsection{Proof of Theorem \ref{thm:perhop-jcats-delay-fifo}}\label{proof:perhop-jcats-delay-fifo}
\begin{figure}[h]
	\centering
	\includegraphics[width= 0.7\linewidth]{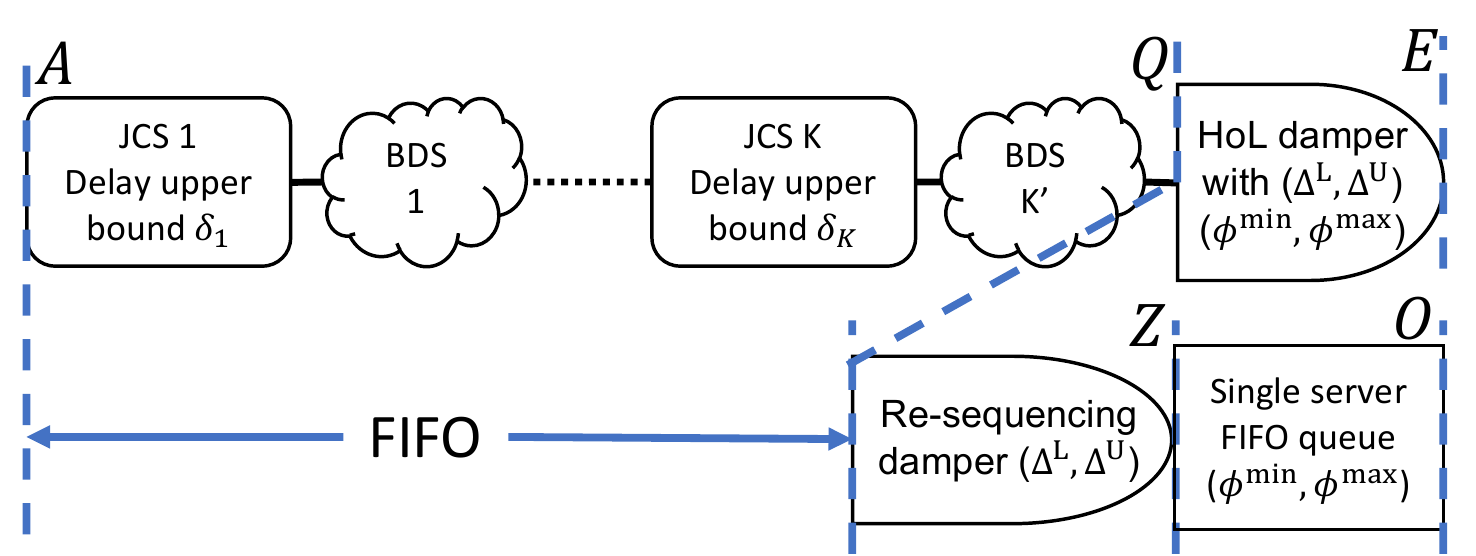}
	\caption{The notations used in the proof of Theorem \ref{thm:perhop-jcats-delay-fifo}.}
	\label{fig:jcats-fifo-proof}
\end{figure}
Consider Fig. \ref{fig:jcats-fifo-proof} where $A$ is the sequence of packet arrival times at JCS $1$ (entrance of the block), $Q$ is the sequence of arrivals to the head-of-line damper, $\tilde{E}$ and $E$ are the sequences of theoretical and actual eligibility times at the head-of-line damper.
By Lemma \ref{lem:jctas-ssq}, the head-of-line damper is equivalent to a re-sequencing damper followed by a single-server FIFO queue with service time in range  $[\phi^{\min},\phi^{\max}]$. Denote with $Z$ the output times of the re-sequencing damper and $O$ as the output of the single-server FIFO queue; the equivalence means that $E=O$.

By Theorem \ref{thm:perhop-fifodamper-delay-fifo}, the delay from $A$ to $Z$ has lower bound $\underline{D}$, upper bound $\overline{D}$ and the jitter bound is $V$. Also, similarly to Corollary \ref{col:concrete-ac-prop}, arrival curve at $Z$ is $\alpha_{\mathrm{reseq}} (t)= \alpha(t+V)$.
% Now, let $\mathcal{P}$ and $\mathcal{P'}$ be the sets of packets seen respectively at $A$ and $Q$, i.e., $\mathcal{P'}\subseteq\mathcal{P}$; by defining $\mathcal{P'}$, we only include the packets of $\mathcal{P}$ that arrive to $Q$; therefore, since the ideal damper is lossless, by \cite[Theorem 4]{mohammadpour2020packet}, the re-sequencing buffer does not increase the worst-case delay and jitter from $A$ to $I$. Hence, the delay and jitter bounds from $A$ to $Z$ are the same as from $A$ to $I$. Moreover, by [\cite{mohammadpour2020packet}, Corollary 1] the propagated arrival curve at $Z$ is $\alpha_{\mathrm{SSQ}}=\alpha_{\mathrm{reseq}} (t+V^0)=\alpha(t+2V^0)$.

 By Lemma \ref{lem:single-server-queue-delay}, the delay upper-bound of the single-server FIFO queue is $\max_{k\in \mathbb{N}}\left\{k\phi^{\max}-\alpha_{\mathrm{reseq}}^{\downarrow}(k) \right\}$ for nonzero processing time, i.e. $\phi^{\max}>0$; otherwise the delay upper-bound is zero. By [\cite{boyer_embedding_2016}, Proposition 7], we obtain $\alpha_{\mathrm{reseq}}^{\downarrow}(k) \geq \alpha^{\downarrow}(k) -V$. Therefore, the delay is upper-bounded by $\theta=\max_{k\in \mathbb{N}}\left\{k\phi^{\max}-\alpha^{\downarrow}(k)+ V \right\}$ for nonzero processing time, i.e., $\phi^{\max}>0$.

 We can see that the delay lower bound is $\phi^{\min}$ as minimum processing time for a packet. This gives the jitter $V_{\mathrm{SSQ}}=\theta-\phi^{\min}$. By summing the bounds from $A$ to $Z$ and the single-server FIFO queue, we obtain the bounds in the statement of the theorem.

\begin{lemma}\label{lem:jctas-ssq}
	Consider a head-of-line damper shown in Fig. \ref{fig:jcats-fifo-proof}. Then, this system can be abstracted as a re-sequencing damper with tolerances ($\tolL,\tolU$) followed by a single-server FIFO queue with service time in range $[\phi^{\min},\phi^{\max}]$.
\end{lemma}

\begin{proof}
	We use the notation in Fig.~\ref{fig:jcats-fifo-proof}. Let $E$ be the sequence of actual eligibility times at the head-of-line damper and $\tilde{E}$ the sequence of theoretical eligibility times. By the discussion that follows \eqref{eq:elig-hol-damper-1} and by Lemma~\ref{lem:hol-definition}, there exist sequences $\bar{E}$ and $\phi$ such that, for every $n$:% \eqref{eq:elig-hol-damper-11} and \eqref{eq:elig-hol-damper-12} for a packet $n$, we have:
	\begin{align}
\nonumber \phi_n&\in [\phi^{\min},\phi^{\max}],\\
	\nonumber \tilde{E}_n-\Delta^{\textrm{L}} \leq &\bar{E}_n \leq \tilde{E}_n+\Delta^{\textrm{U}},\\
	E_n & = \max(\bar{E}_n, E_{n-1})+\phi_n.\label{eq:lemHoL1acx}
	\end{align}
	Let us construct a re-sequencing damper with tolerances ($\tolL,\tolU$) and sequence of actual eligibility times $Z$ such that
	\begin{align}\label{eq:hol-damper-lemma-proof-1}
	\nonumber \tilde{E}_n-\Delta^{\textrm{L}} &\leq \bar{E}_n \leq \tilde{E}_n+\Delta^{\textrm{U}}, \\
	Z_1 = \bar{E}_1,~~&Z_n = \max\left\{\bar{E}_n, Z_{n-1}\right\}.
	\end{align}
	Now consider a  single-server FIFO queue with sequence of service times equal to $\phi$ and input sequence $Z$. Then, the output sequence from the FIFO queue, $O$, is
	\begin{align}\label{eq:lemHoL1aa}
	O_1 = Z_1 +\phi_1, O_n = \max(O_{n-1},Z_n)+\phi_n; n\geq 2.
	\end{align}
We now show by induction that $O_n=E_n$ for every $n \geq 1$. This holds for $n=1$. Assume that it holds for $n-1$. Observe first that $Z_{n-1}\leq O_{n-1}$ (because $\phi_n\geq 0$) and therefore, by the induction hypothesis,
\begin{equation}\label{eq:lemHoL1ad}
  Z_{n-1}\leq E_{n-1}.
\end{equation}
By \eqref{eq:lemHoL1aa} and again the induction hypothesis:
\begin{equation}
  O_{n}=\max(E_{n-1}, Z_n)+\phi_n.
  \label{eq:lemHoL1ab}
\end{equation}
By \eqref{eq:hol-damper-lemma-proof-1}:
\begin{align}\label{eq:lemHoL1ac}
   O_{n}&=\max(E_{n-1}, \bar{E}_n, Z_{n-1})+\phi_n\\
    & =\max(E_{n-1}, \bar{E}_n)+\phi_n,
\end{align}
because of \eqref{eq:lemHoL1ad}. It follows from \eqref{eq:lemHoL1acx} that $O_n=E_n$.
\end{proof}

\begin{lemma}\label{lem:single-server-queue-delay}
	Consider a flow with per-packet arrival curve $\alpha$ that enters a single-server FIFO queue. Suppose that a head of line packet $n$ has a processing time $\phi_n\in\left[\phi^{\min},\phi^{\max}\right]$. Then a delay bound of the flow $\theta$ is:
	\begin{align}
	\theta = \max_{i\in \mathbb{N}}\left\{i\phi^{\max}-\alpha^{\downarrow}(i) \right\}.
	\end{align}
\end{lemma}
\begin{proof}
	Let us call $I_n$ and $O_n$ as arrival and departure times of packet $n$.
	%	Then we have:
	%	\begin{align}
	%	O_n - I_n \geq \phi_n \geq \phi^{\min},
	%	\end{align}
	%	that proves the lower bound.
	By Lemma \ref{lem:single-server-queue-inout}, we have:
	\begin{align}
	O_n = \max_{m\leq n}\left\{I_m+\sum_{i=m}^{n}\phi_i\right\}.
	\end{align}
	We subtract $I_n$ from both sides:
	\begin{align}
	\nonumber O_n - I_n &= \max_{m\leq n}\left\{I_m+\sum_{i=m}^{n}\phi_i\right\}-I_n\\
	&=\max_{m\leq n}\left\{\sum_{i=m}^{n}\phi_i -\left(I_n-I_m\right)\right\}.
	\end{align}
	By \cite[Section III.E]{mohammadpour2020packet}, we have $I_n-I_m \geq \alpha^{\downarrow}(n-m+1)$; therefore,
	\begin{align}
	O_n - I_n \leq \max_{m\leq n}\left\{\sum_{i=m}^{n}\phi_i -\alpha^{\downarrow}(n-m+1)\right\}.
	\end{align}
	Since $\phi_i \leq \phi^{\max}$:
	\begin{align}
	\nonumber O_n - I_n &\leq \max_{m\leq n}\left\{(m-n+1)\phi^{\max} -\alpha^{\downarrow}(n-m+1)\right\}\\
	&\leq \max_{i\in \mathbb{N}}\left\{i\phi^{\max}-\alpha^{\downarrow}(i) \right\}=\theta.
	\end{align}
\end{proof}

\begin{lemma}\label{lem:single-server-queue-inout}
	Consider a single-server FIFO queue. A packet $n$ arrives at time $I_n$ and the service time is $\phi_n$ when it is at the head of the queue. Then, the departure time of packet $n$ is $O_n$ and computed as:
	\begin{align}
	O_n = \max_{m\leq n}\left\{I_m+\sum_{i=m}^{n}\phi_i\right\}.
	\end{align}
\end{lemma}
%\jylb{This lemma is the same as Lemma 2}
\begin{proof}
	Since we have for a single-server FIFO queue:
	\begin{align}\label{eq:ssq-inout-proof-1}
	O_1=I_1+\phi_1,~~O_n = \max\left(I_n,O_{n-1}\right)+\phi_n,
	\end{align}
	by Lemma \ref{lem:rec-to-closed} the statement is proven.
%	We prove the lemma by induction. Base case $n=1$.
%	\begin{align}
%	O_1=I_1+\phi_1,
%	\end{align}
%	as required by the lemma.
%	
%	Induction step. We assume that the lemma holds for all $i<n$. Then for packet $n-1$, we have:
%	\begin{align}\label{eq:ssq-inout-proof-n-1}
%	O_{n-1} = \max_{m\leq n-1}\left\{I_m+\sum_{i=m}^{n-1}\phi_i\right\}.
%	\end{align}
%	Then, by \eqref{eq:ssq-inout-proof-1} we have:
%	\begin{align}
%	\nonumber O_n &= \max\left(I_n,O_{n-1}\right)+\phi_n\\
%	\nonumber &=\max\left(I_n+\phi_n,\max_{m\leq n-1}\left\{I_m+\sum_{i=m}^{n-1}\phi_i\right\}+\phi_n\right)\\
%	\nonumber &=\max\left(I_n+\phi_n,\max_{m\leq n-1}\left\{I_m+\sum_{i=m}^{n}\phi_i\right\}\right)\\
%	&=\max_{m\leq n}\left\{I_m+\sum_{i=m}^{n}\phi_i\right\},
%	\end{align}
%	that completes the proof.
\end{proof}

\subsection{Proof of Theorem \ref{thm:perhop-fifodamper-delay-nonfifo}} \label{proof:perhop-fifodamper-delay-nonfifo}
\begin{figure}[h]
	\centering
	\includegraphics[width=0.9 \linewidth]{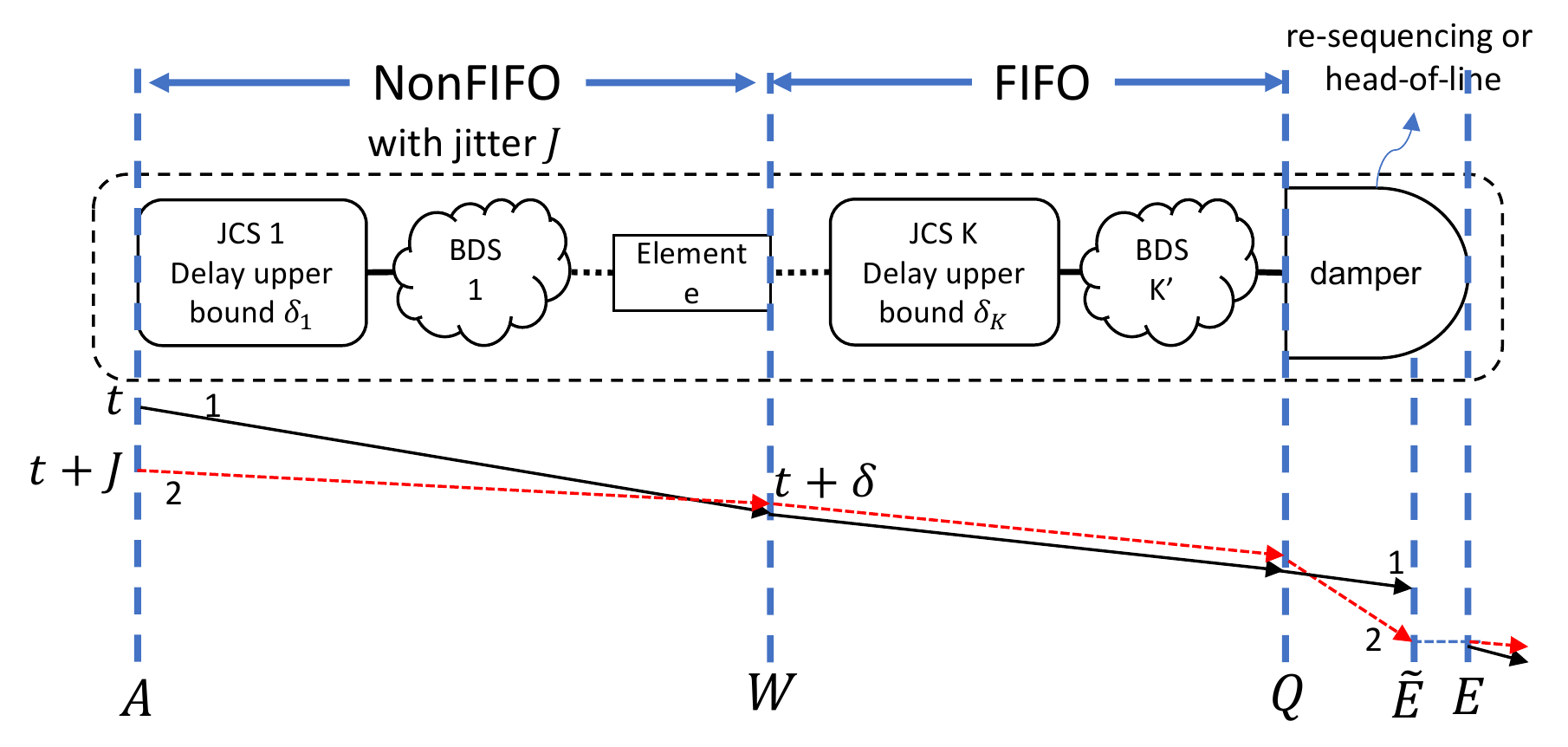}
	\caption{The notations used in the proof of Theorem \ref{thm:perhop-fifodamper-delay-nonfifo}.}
	\label{fig:fifodamper-nonfifo-reseq}
\end{figure}

Consider Fig. \ref{fig:fifodamper-nonfifo-reseq} where $A$ is used to denote the arrival times to JCS $1$, $W$ the departure times from element $e$, $Q$ the arrival times to the damper, $\tilde{E}$ and $E$ respectively the theoretical and actual eligibility times at the damper. Now for packets $m,n: m<n$ , since from output of element $e$ to the input of the damper is FIFO, we have $W_m \leq W_n$. Then, as the jitter from JCS $1$ to element $e$ is bounded by $J$, we have:
\begin{align}\label{eq:fifodamper-nonfifo-proof-1}
\nonumber &(W_{n}-A_{n}) - (W_{m}-A_{m}) \leq J\\
\nonumber \mbox{thus }&A_{m} - A_{n} \leq J+W_{m}-W_{n} \leq J\\
\mbox{thus }& A_{n} \geq  A_{m} - J.
\end{align}
%Note that the propagation delay is the same for two packets. % and imposes zero jitter.
%Now, we analyze the delay bounds for the case of re-sequencing dampers and head-of-line dampers separately.
%If the damper is a re-sequencing damper ($\Delta^{\mathrm{L}},\Delta^{\mathrm{U}}$),
Now by \eqref{eq:elig-fifo-damper}, we have:
\begin{align}\label{eq:fifodamper-nonfifo-proof-2}
\nonumber \tilde{E}_n-\Delta^{\textrm{L}} &\leq \bar{E}_n \leq \tilde{E}_n+\Delta^{\textrm{U}}, \\
E_1 = \bar{E}_1,~~&E_n = \max\left\{\bar{E}_n, E_{n-1}\right\}.
\end{align}
By definition, the re-sequencing damper behaves as a damper with tolerances ($\Delta^{\mathrm{L}},\Delta^{\mathrm{U}}$) followed by a re-sequencing buffer, where $\bar{E}$ is the output of the damper with tolerance and $E$ is the output of the re-sequencing buffer.
Now, let $n$ be fixed and define packet index $u$ as
\begin{align}
u=\max\left\{k\leq n| \bar{E}_k = \max_{j\leq n}\left\{\bar{E}_j\right\}\right\}.
\end{align}
Then, we have $E_n - A_n = \bar{E}_u -A_n$.
Combining with \eqref{eq:fifodamper-nonfifo-proof-1} for packets $u$ and $n$, we obtain:
\begin{align}
E_{n} -A_{n}&\leq \bar{E}_{u}-A_{u}+J.
\end{align}
Since $\bar{E}$ is the output of the damper with tolerance, by Theorem \ref{thm:perhop-concrete-delay}, $\bar{E}_{u}-A_{u}\leq \overline{D}$; hence $E_{n} -A_{n}\leq \overline{D}+J$, which proves the delay upper bound.
By \eqref{eq:fifodamper-nonfifo-proof-2} and Theorem \ref{thm:perhop-concrete-delay}, we have:
\begin{align}
E_{n} -A_{n} \geq \bar{E}_{n}-A_{n}\geq \underline{D},
\end{align}
which proves the delay lower bound for this case. Since the delay lower-bound is not changed, and the upper-bound is increased by $J$, hence the jitter bound is increased by $J$.

\textbf{Proof of tightness.}
The tightness scenario for delay lower-bound is exactly the same as tightness scenario in Theorem \ref{thm:perhop-concrete-delay} where a single packet in isolation reaches the delay lower-bound $\underline{D}$.

For the delay upper-bound and jitter bound tightness, consider two packets $1$ and $2$ as shown in Fig \ref{fig:fifodamper-nonfifo-reseq} that arrive at times $t$ and $t+J$ in TAI. Assume every local clock (of JCS or damper) $\mathcal{H}_i$ is adversarial and faster than TAI such that for any delay measurement $d$, we have:
\begin{align}
\nonumber d^{\mathcal{H}_{\mathrm{TAI}}} &= \min\left(\rho d^{\mathcal{H}_i}+\eta,d^{\mathcal{H}_i}+2\omega\right).
\end{align}
%
%\jylb{How about $\omega$~?}
Let us call the worst-case delay, in TAI, from input of JCS $1$ to the output of $e$, as $\delta$. Suppose that packets $1$ and $2$ experience delays of $\delta$ and $\delta-J$, in TAI, to leave element $e$, i.e., $W_1 =W_2=t+\delta$, and packet $2$ arrives just before packet $1$. Also, both experience the same delay from output of element $e$ to the input of the damper while packet $2$ is still prior to packet $1$ due to the FIFO assumption. Since packet $1$ arrives after packet $2$, it is released after packet $2$ becomes eligible (even if its theoretical eligibility time has passed).

Now, assume packet $2$ experiences a delay, in TAI, equal to $\overline{D}$ from $A$ to $E$, i.e., $E_2 = A_2 + \overline{D} = t+J+\overline{D}$ (the packet that reaches the upper-bound in tightness scenario of Theorem \ref{thm:perhop-concrete-delay}). Therefore, packet $1$ is released after packet $2$, i.e., $E_1=E_2=t+J+\overline{D}$.
Hence, the delay of packet $1$ from $A$ to $E$ is:
\begin{align}
E_1 - A_1 = (t+J+\overline{D}) - (t) = J+\overline{D},
\end{align}
that is equal to the delay upper-bound of the theorem statement.

Now, we verify the assumptions;

\noindent 1) The jitter from JCS $1$ to the output of $e$ is not larger that $J$: The delay of packet $1$ and packet $2$ are respectively $\delta$ and $\delta-J$ in TAI, and therefore the jitter is $J$.

\noindent 2) The FIFO constraint of the damper is not violated: Packet $2$ arrives before packet $1$, $Q_2\leq Q_1$, and also leaves before is $E_2\leq E_1$.

Hence, we showed an execution trace that with packet $2$ experiencing a delay of   $\overline{D}$, packet $1$ experiences a delay of $\overline{D}+J$.
Since there execution traces where in one, a packet reaches the lower-bound of Theorem \ref{thm:perhop-concrete-delay} and in another one, a packet reaches the delay upper-bound of Theorem \ref{thm:perhop-concrete-delay} plus $J$, we have that the jitter Theorem \ref{thm:perhop-concrete-delay} is increased by $J$.

\subsection{Proof of Theorem \ref{thm:perhop-jcats-delay-nonfifo}}\label{proof:perhop-jcats-delay-nonfifo}
Consider Fig. \ref{fig:fifodamper-nonfifo-reseq} where $A$ is used to denote the sequence of arrival times to JCS $1$, $W$ the departure times from element $e$, $Q$ the arrival times to the damper, $\tilde{E}$ and $E$ respectively the theoretical and actual eligibility times at the damper. By Lemma \ref{lem:jctas-ssq}, we abstract it as a re-sequencing damper followed by a single-server FIFO queue. Let $Z$ denote the sequence of departure times from the re-sequencing damper; then, by Theorem \ref{thm:perhop-fifodamper-delay-nonfifo}, for a packet $n$, we have:
\begin{align}\label{eq:hol-nonfifo-proof-2}
\underline{D} \leq Z_n - A_n \leq \overline{D}+J,
\end{align}
and the jitter from $A$ to $Z$ is $V+J$. As a result, by \cite[Lemma 1]{mohammadpour2020packet}, an arrival curve at the output of the re-sequencing damper (input of the single-server FIFO queue) is $\alpha_{\mathrm{reseq}}=\alpha(t+V+J)$. Then by Lemma \ref{lem:single-server-queue-delay} for nonzero processing time:
\begin{align}\label{eq:hol-nonfifo-proof-3}
\nonumber E_n - Z_n &\leq \max_{k\in \mathbb{N}}\left\{k\phi^{\max}-\alpha_{\mathrm{reseq}}^{\downarrow}(k)\right\}\\
& \leq \max_{k\in \mathbb{N}}\left\{k\phi^{\max}-\alpha^{\downarrow}(k)+ V+J \right\} = \theta+J,
\end{align}
where $\theta$ is defined in \eqref{eq:theta}. Finally, by \eqref{eq:hol-nonfifo-proof-2}, we have,
\begin{align}
\nonumber E_n-A_n &= (E_n-Z_n) + (Z_n-A_n) \\
&\leq \overline{D}+  J + (\theta +J)1_{\{\phi^{\max}>0\}},
\end{align}
which proves the delay upper bound. Note that by Theorem \ref{thm:perhop-jcats-delay-fifo}, an upper-bound on the delay is $\overline{D}+ \theta 1_{\{\phi^{\max}>0\}}$.

Using minimum processing time and \eqref{eq:hol-nonfifo-proof-2}, we have:
\begin{align}
\nonumber  E_{n} -A_{n} &= (E_n-Z_n) + (Z_n-A_n)  \\
&\geq \phi^{\min} +\underline{D},
\end{align}
which proves the delay lower bound. Since the delay lower-bound is not changed, and the upper-bound is increased by $J+J 1_{\{\phi^{\max}>0\}}$, the jitter bound is increased by $J+J 1_{\{\phi^{\max}>0\}}$.

%Case HoL damper.
%\todo{assumption: 1)$\alpha$ is continuous. 2) $\alpha^+(0)=1$}
%
%Consider $n$ packets, where $n=\alpha(V+J-\varepsilon)$ for some $\varepsilon>0$. We set the sequence of their arrival times to the block, $A$, such that:
%\begin{align}
%A_i = \alpha^{\downarrow}(i) ~~~\forall i=1,\dots,n.
%\end{align}
%Note that $A_1=0$ as $\alpha^+(0)=1$; $A_n=\alpha^{\downarrow}(n)=J+V-\varepsilon$, as $n=\alpha(J+V)$ and by continuity of $\alpha$, we have $\alpha^{\downarrow}(n) = V+J$. Let us define packet $m$ such that $m=\alpha(V)$; then $A_m = \alpha^{\downarrow}(m)=V$.
%
%Assume every local clock (of JCS or damper) $\mathcal{H}_i$ is adversarial and faster than TAI such that for any delay measurement $d$, we have:
%\begin{align}
%\nonumber d^{\mathcal{H}_{\mathrm{TAI}}} &= \rho d^{\mathcal{H}_i}+\eta.
%\end{align}
%
%Let us call the worst-case delay, in TAI, from input of JCS $1$ to the output of $e$, as $\delta$. For the departure times from $e$ in TAI, we set:
%\begin{align}
%\nonumber W_i&=A_i+\delta~~~~\forall i=1,\dots,m-1,\\
%\nonumber W_m&=A_m+\delta=V+\delta,\\
%\nonumber W_i&=W_m=V+\delta ~~~~ \forall i=m+1,\dots,n-1,\\
%W_n&=A_n+\delta-J=\delta+V-\varepsilon
%\end{align}
%
%By this construction, we have $W_1\leq W_2\leq \dots \leq W_{m-1}$ $W_n < W_m=W_{m+1}=\dots=W_{n-1}$.
%We set the packet arrivals to the entrance of damper as: 

\end{document}